\documentclass[a4paper,11pt]{article}
\pdfoutput=1 

\usepackage{jcappub} 

\usepackage[T1]{fontenc} 
\usepackage{hyperref}
\usepackage[dvipsnames, table]{xcolor}

\usepackage{cleveref}

\usepackage{slashed}
\usepackage{mathtools}
\usepackage{textcomp}
\usepackage{physics}
\usepackage{soul}

\usepackage{parskip}

\usepackage{feynmp-auto}

\usepackage{tcolorbox}

\usepackage{graphicx}  
\usepackage{booktabs}  
\usepackage{array}     
\usepackage{tabularx} 
\usepackage {subfigure} 
\usepackage{amsmath,mathrsfs,amsfonts,amssymb}

\newcommand{\be}{\begin{equation}}
\newcommand{\ee}{\end{equation}}
\newcommand{\bea}{\begin{eqnarray}}
\newcommand{\eea}{\end{eqnarray}}
\definecolor{lightblue}{rgb}{0.8,0.85,1}

\usepackage{tikz}
\usepackage{pgfplots}
\pgfplotsset{compat=1.18}
\usetikzlibrary{3d,patterns,arrows.meta}
\usepackage{amssymb}

\title{Quantum Gravity Meets DESI: Dynamical Dark Energy in Light of the Trans-Planckian Censorship Conjecture}

\author[a, b,\dagger]{Chunyu Li,}
\author[c,d,\dagger]{Junkai Wang,\note[\textdagger]{Co-first author.}}
\author[a, b]{Dongdong Zhang,}
\author[e,b,f]{Emmanuel N. Saridakis,}
\author[a, b]{Yi-Fu Cai}

\affiliation[a]{Department of Astronomy, School of Physical Sciences, University of Science and Technology of China, Hefei 230026, China}

\affiliation[b]{CAS Key Laboratory for Research in Galaxies and Cosmology, School of Astronomy and Space Science, University of Science and Technology of China, Hefei 230026, China}

\affiliation[c]{School of Physics, Nanjing University, Nanjing 210023, China}
\affiliation[d]{Yau Mathematical Sciences Center, Tsinghua University, Beijing 100084, China}
\affiliation[e]{National Observatory of Athens, Lofos Nymfon 11852, Greece}
\affiliation[f]{Departamento de Matem\'{a}ticas, Universidad Cat\'{o}lica del Norte, Avda. Angamos 0610, Casilla 1280, Antofagasta, Chile}

\emailAdd{springrain@mail.ustc.edu.cn}
\emailAdd{junkaiwang@smail.nju.edu.cn}
\emailAdd{don@mail.ustc.edu.cn}
\emailAdd{msaridak@noa.gr}
\emailAdd{yifucai@ustc.edu.cn}

\abstract{Recent DESI DR2  observations indicate that dark energy has crossed from  phantom     to   quintessence regime, a behavior known as the quintom-B realization. In this work  we constrain dynamical dark energy and modified gravity  using the swampland Trans-Planckian Censorship Conjecture (TCC), which forbids eternal acceleration since in this case  any trans-Planckian quantum fluctuation would eventually stretch beyond the Hubble radius, breaking the applicability of any effective field theory and cosmological techniques. By combining DESI DR2 data with the TCC criterion, we impose tight constraints on the dark energy equation of state and its parameter space in scenarios such as the Chevallier-Polarski-Linder, Barboza-Alcaniz, Jassal-Bagla-Padmanabhan, EXP and LOG parameterizations,  significantly constraining the quintom-A behavior. Also we examine models within  the framework of $f(T)$ and $f(Q)$ modified gravity theories, demonstrating that TCC is very powerful to constrain or exclude  them, a result that indicates   the necessity to consider infrared  modifications on General Relativity apart from the usual ultraviolet ones. Our findings imply that viable dynamical dark energy scenarios must asymptotically transit to deceleration, shedding light  on new physics consistent with both cosmological observations and quantum gravity principles.} 

\keywords{Swampland, TCC Criterion, Dynamical Dark Energy, Modified Gravity, DESI }

\begin{document}
\maketitle
\flushbottom

\section{Introduction}

The accelerated expansion of the universe was first observed in 1998 through studies of Type Ia supernovae  \cite{SupernovaSearchTeam:1998fmf,SupernovaCosmologyProject:1998vns}. These observations motivated the proposal of dark energy and the subsequent development of the \(\Lambda\)CDM paradigm. In this basic scenario dark energy is represented by the cosmological constant \(\Lambda\), which uniformly permeates space and accounts for approximately 70\% of the total energy density of the universe.

Recent observations from the Dark Energy Spectroscopic Instrument (DESI) have indicated that dark energy may evolve and weaken over time  \cite{DESI:2024mwx,DESI:2025zgx,DESI:2025fii,DESI:2025ejh,Gongbo2025}, thereby challenging its assumed constancy within the \(\Lambda\)CDM scenario. Many proposals have been put forward trying to account for time-varying dark energy \cite{Yin2025}.
If confirmed, this time-varying behavior could significantly reshape our understanding of the universe's expansion history and its ultimate fate.
In particular, DESI analyses indicate that the dark energy equation of state (EoS) parameter may cross from $w < -1$ to $w > -1$ over time---a phenomenon known as the quintom-B behavior  \cite{Feng:2004ad,Huterer:2004ch,Cai:2009zp}.
For the conventional scalar field models for dark energy, this scenario violates the Null Energy Condition (NEC), thereby necessitating non-trivial modifications to the \(\Lambda\)CDM scenario. 
Hence, several theoretical constructions, such as modified gravity  \cite{Wetterich:1994bg,Amendola:1999er,Saridakis:2012jy,CANTATA:2021asi,Yang:2024kdo,Yang:2025kgc,Yang:2025mws}, interacting dark energy  \cite{Giare:2024smz,Li:2024qso,Zhai:2025hfi,Pan:2025qwy}, non-minimal coupled gravity  \cite{Ye:2024ywg, Pan:2025psn} etc,  demonstrating that the quintom scenario can be achieved while simultaneously preserving the NEC.

However, the observed weakening in the dark energy equation of state is not unexpected from the perspective of quantum gravity (QG) \cite{Cai:2019igo,Payeur:2024kyy,Arjona:2024dsr,Bhattacharya:2024kxp,Brandenberger:2025hof,Anchordoqui:2025fgz}. For example, the Trans-Planckian Censorship Conjecture (TCC) posits that trans-Planckian quantum fluctuations should not propagate across the Planck scale, since this would imply that unknown trans-Planckian physics would affect the  low-energy behavior of our theories, or in other words that our effective field theories  (like general relativity plus quantum fields)   that we use to describe the Universe evolution would not be  ``effective field theories''   compatible with an ultraviolet-complete theory of quantum gravity
 \cite{Martin:2000xs,Rob2000wr,Rob2012}.

Extensive earlier studies have employed  the TCC to constrain inflation and primordial black holes in the early universe, however it can also be applied to constrain the late-time universe behavior, such as various quintessence scalar-field models  \cite{Agrawal:2018own,  Heisenberg:2018yae, Li:2019ipk, Cicoli:2020cfj, Payeur:2024kyy,Anchordoqui:2025fgz}.
Although applying the TCC criterion to the very early universe—particularly during the inflationary epoch—remains controversial  due to uncertainties surrounding the driving force of inflation, or the nature of non-perturbative quantum gravity correction,  it is crucial to note that in the post-inflationary cosmic history the energy scale is much lower than the Planck scale, making these controversial issues to disappear. 

The TCC criterion implies that within the framework of effective field theory of quantum gravity,  any cosmological model that predicts a perpetually accelerating expansion in the future is necessarily excluded, since this would imply that sub-planckian small-scale quantum fluctuations will eventually  be stretched beyond the horizon and be classicalized.
Nevertheless, previous applications of the TCC did not fully incorporate it at the data level, potentially leading to inconsistencies with quantum gravity or necessitating the imposition of stronger constraints through combined analyses.

In this work, we elucidate the joint constraints imposed by the quintom-B scenario and the TCC for the first time. Our article is organized as follows. In Section \ref{sec_2}  we review the historical evolution of dynamical dark energy research along with recent advancements, including novel insights from DESI DR2 data \cite{DESI:2025zgx}. Subsequently, we introduce the TCC criterion and we discuss its implications for dynamical dark energy models. In Section \ref{sec_3}  we perform a combined analysis of quintom-B and   TCC constraints, using well-known parameterizations such as the Chevallier-Polarski-Linder (CPL) and the Barboza–Alcaniz (BA) ones, analyzing the allowed parameter space. Moreover, in Section \ref{sec_4} we examine two modified gravity models based on the $f(T),f(Q)$ framework, and by applying joint constraints  we eventually rule them out, thereby demonstrating the robust limitations imposed on new physical models. Finally, we conclude our work in Section \ref{sec_5}.

\textbf{Note Added:} While this work was at its final stage, a work by Brandenberger \cite{Brandenberger:2025hof} appeared, demonstrating that DESI results are consistent with the expectation form the TCC criterion. 
In our study, we investigated the quintom-B behavior suggested by  DESI DR2 data and we clarified the TCC constraints on various parameterizations of the equation of state \(w(a)\) and, for the first time, we applied the TCC to modified gravity theories under the light of recent observational datasets.

\section{Quantum Gravity Meets Dynamical Dark Energy}
\label{sec_2}

In this section we first present the basic features of dynamical dark energy, and then we review the  TCC Criterion, applying it to the simple cosmological constant as well as to dynamical dark energy scenarios.

\subsection{Historical Evolution of Dynamical Dark Energy Scenarios}

In modern cosmology, observational evidence from Type Ia supernovae led to the accelerating cosmic expansion, thereby establishing dark energy as a fundamental component of the universe. Within the standard $\Lambda$CDM paradigm, dark energy is modeled by a simple cosmological constant, with the EoS $w= 
P/\rho\equiv -1$. Nevertheless, despite its empirical successes, $\Lambda$CDM encounters theoretical and observational challenges and tensions \cite{DiValentino:2025sru}, which have motivated the investigation of dynamical dark energy \cite{Copeland:2006wr}.

In order to construct dynamical dark energy scenarios, one typically adopts specific parameterizations of the dark-energy equation-of-state parameter. For instance, the $w$CDM model assumes a constant $w$ that deviates from $-1$, whereas the $w_0w_a$CDM parameterization describes $w$ as evolving over time with two free parameters. When applying the Chevallier-Polarski-Linder (CPL) parameterization  $w(a)=w_0+w_a(1-a)$  \cite{Chevallier:2000qy,Linder:2002et}, Planck 2018 \cite{Planck:2018vyg} constrains both $w_0$ and $w_a$ to values appearing inconsistency with $\Lambda$CDM model at approximately   $2\sigma$ confidence level, thereby favoring a phantom scenario. Moreover, recent baryon acoustic oscillation (BAO) measurements from DESI  \cite{DESI:2024mwx}, when combined with CMB and supernova data, have provided evidence at significance levels of $2.5\sigma$, $3.5\sigma$, and $3.9\sigma$ from the PantheonPlus, Union3, and DESY5 datasets, respectively. Intriguingly, the DESI data suggest a quintom-B behavior, and subsequent results from the DESI Full-Shape analysis  \cite{DESI:2024hhd} and DESI DR2  \cite{DESI:2025zgx,Gu:2025xie} further consolidate this trend.

Dynamical dark energy models allow the dark energy density to vary over time, implying that the EoS parameter may depart from $-1$. Typically, models with $w\geq-1$ are classified as quintessence scenario, whereas those with $w\leq-1$ as phantom scenario\footnote{The terms ``quintessence'' and ``phantom'' are also used to refer to scalar field $(\varphi)$ realizations  of dark energy; however in our context  these terms solely denote the phases \(w > -1\) and \(w < -1\), not demanding a particular realization of dynamical dark energy scenarios, making the discussion more general and applicable.}. When $w$ crosses the $-1$ boundary, the model is termed as quintom scenario. A variety of scalar field models have been proposed to elucidate the dynamical behavior of dark energy, including quintessence  \cite{Ratra:1987rm}, phantom  \cite{Caldwell:1999ew,Caldwell:2003vq,Cline:2003gs}, k-essence  \cite{Chiba:1999ka,Armendariz-Picon:2000nqq}, etc. However, the ``No-Go'' theorem\cite{Vikman:2004dc,Xia:2007km}, which forbids a single perfect fluid or a single scalar field from crossing the phantom divide, and the requirement of preserving the null energy condition (NEC) each place constraints on model building, motivating the exploration of alternative approaches. Consequently, developments in modified gravity \cite{Wetterich:1994bg,Amendola:1999er,Saridakis:2012jy,CANTATA:2021asi,Yang:2024kdo,Yang:2025kgc}, interacting dark energy \cite{Giare:2024smz,Li:2024qso,Zhai:2025hfi,Pan:2025qwy} and non-minimally coupled gravity \cite{Ye:2024ywg, Pan:2025psn} have been advanced to address the evolution of the dark sector in a self-consistent manner.

\subsection{General Constraint from TCC Criterion}
 
\begin{figure*}[t]
\centering
\includegraphics[width=1\textwidth]{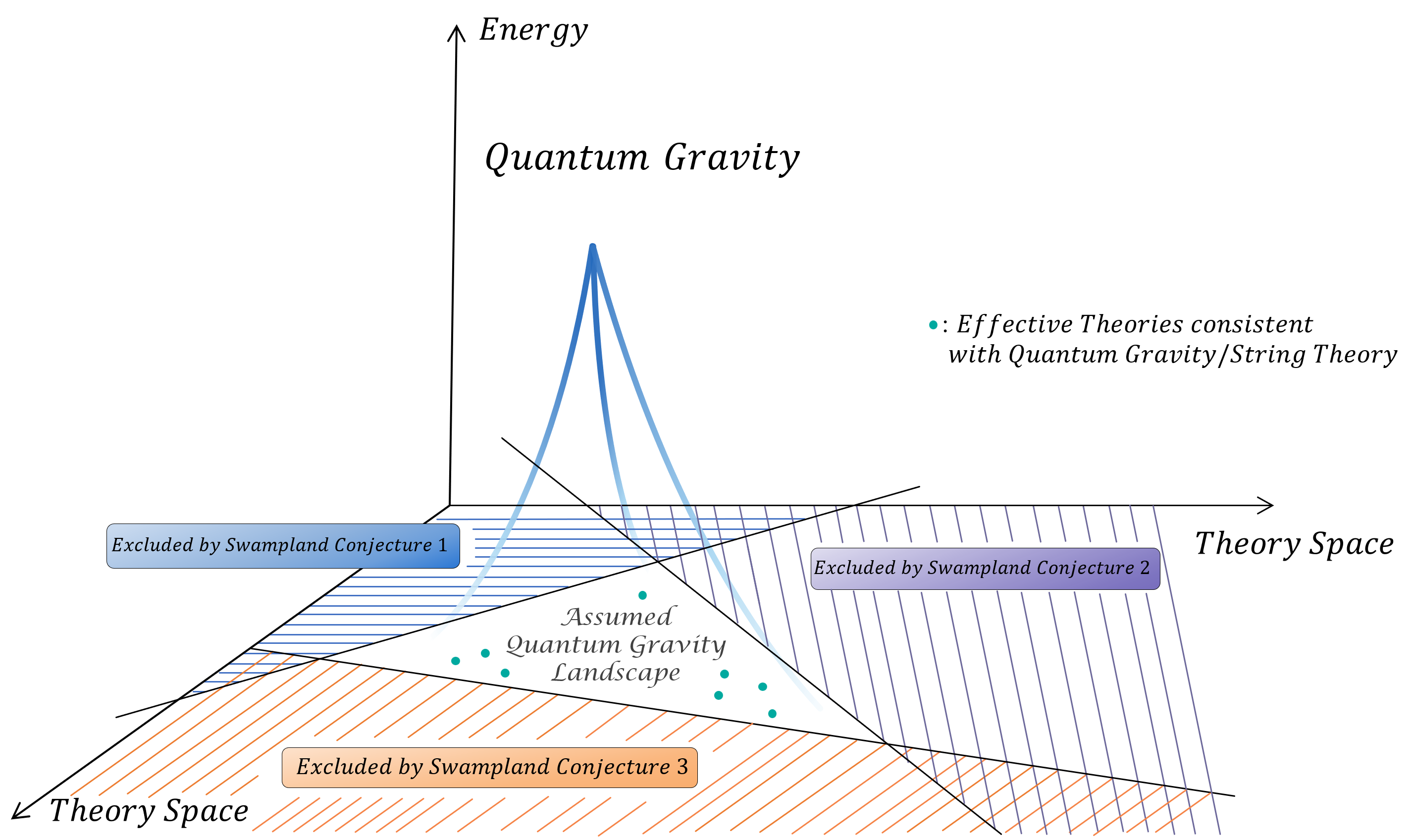}
\caption{Demonstration of Landscape \& Swampland in Quantum Gravity.  Conjectures 1,2,3 come from several swamplandish considerations \cite{vanBeest2021,Agmon2022}.}
\label{swamp}
\end{figure*}

The Trans-Planckian Censorship Conjecture    \cite{Martin:2000xs} is one of the components of Swampland Conjectures, first proposed in 2005 by Vafa  \cite{Vafa:2005ui}, aiming to give the consistent conditions of effective field theory with quantum gravity, excluding incompatible effective theories. 
The Swampland Conjecture Project  \cite{vanBeest2021,Agmon2022} is motivated and confirmed by our knowledge of quantum gravity obtained from perturbative string theory, string dualities, black hole physics, holographic principle, etc. Effective Field Theories (EFTs) that satisfy these constraints are believed to lie within the quantum gravity ``landscape'', whereas those that violate them are conjectured to belong to the ``swampland'', as shown in  Fig.~\ref{swamp}. Furthermore, there exist a  speculation that the UV-complete quantum gravity theory is  unique \cite{vafa2020}, giving out all possible EFTs from the landscape of a single QG theory.

The TCC criterion is rooted in our understanding of gravitational effective field theories, which capture the low-energy behavior of a theory by integrating out high-energy degrees of freedom, as in the four-Fermi and pion theories. Due to our limited knowledge of quantum gravity at the Planck scale, we expect gravity to have a gravitational EFT description. However, cosmic expansion redshifts high-frequency modes into the infrared, potentially causing UV-IR mixing, which can render the Hilbert space ill-defined and compromise unitarity.

More precisely, when sub-Planckian modes are redshifted and eventually stretch beyond the Hubble radius during cosmic expansion, they undergo a classicalization process \cite{Rob1990,Kiefer2008}. In the framework of EFT, this implies that if any perturbation observed today on superhorizon scales were to originate from sub‑Planckian scales, it would signal a breakdown of the unitary EFT description. In other words, a consistent gravitational EFT should forbid any sub-Planckian degree of freedom to become classicalized. In summary, this condition imposes a bound on the evolution of the scale factor and the Hubble parameter. 

\textbf{TCC Criterion:}  \textit{After the initial time $t_i$, when the gravitational EFT description becomes valid, the scale factor $a(t)$ and Hubble parameter $H(t)$ should satisfy }

\be
\frac{a\left(t_f\right)}{a\left(t_i\right)}  <C\cdot\frac{r_H}{\ell_{p l}}=\frac{C\cdot M_{pl}}{H\left(t_f\right)},
\ee

\textit{for any final time \(t_f > t_i\).  Equivalently, the e-fold number $\Delta N=\ln\frac{a(t_f)}{a(t_i)}$ should satisfy}

\be
\Delta N \equiv \int_{t_i}^{t_f} H\, dt  <  \ln\frac{C\cdot M_{pl}}{H(t_f)},
\ee

\textit{for any final time \(t_f > t_i\).}

Here we work   in natural units. $r_H=\frac{1}{H}$ denotes the Hubble radius, and quantities \( \ell_{pl}\) \& \(M_{pl}\) denote respectively the Planck length and Planck mass, obeying $\ell_{ {pl}} = \frac{1}{M_{ {pl}}}$. The constant \(C\) is a dimensionless parameter of order unity, i.e., \(C \sim \mathcal{O}(1)\). For convenience without loss of generality, we shall set \(C = 1\) in the following discussion of the TCC criterion.

By differentiating the TCC criterion with respect to $t_f$, one obtains the differentiated version of the TCC criterion, namely
\be
H^2(t_f) < -\frac{d}{dt}H(t_f),
\ee
which is equivalent to
\be
\left.\frac{d^2}{dt^2}a(t)\right|_{t_f}< 0.
\ee
The differentiated version is in general different from and much stronger than the original one (unless certain concavity properties are required). If the differentiated version of the TCC holds, then the original TCC is automatically satisfied.  That is, if  TCC is satisfied at time \(t_b\) and   \(\ddot{a} < 0\) after time $t_b$, then we have 
\be
\Delta N \equiv \int_{t_a}^{t_c} H\, dt \equiv \int_{t_a}^{t_b} H\, dt + \int_{t_b}^{t_c} H\, dt < \ln\frac{M_{pl}}{H(t_b)} + \ln\frac{H(t_b)}{H(t_c)} = \ln\frac{M_{pl}}{H(t_c)}.
\label{dTCC}
\ee
Notably, the differentiated version of the TCC does not depend on the value of \(M_{{pl}}\); indeed, Eq.~\ref{dTCC} reveals that $M_{pl}$ serves as the initial conditions. This reflects that the TCC criterion is automatically satisfied in a decelerating universe, regardless of the details of its evolution, as long as the initial conditions saturate the TCC criterion. 

However, since our universe is currently undergoing accelerated expansion, the TCC is not automatically fulfilled; rather, it imposes specific constraints on the present cosmological evolution. On the other hand, as $t \to +\infty$, according to the differentiated version of the TCC, $w\ge -1/3$ must eventually be satisfied, leading to a decelerating universe. Consequently, the TCC naturally limits the evolution of an accelerating universe until the acceleration stops.

\subsection{Implications of TCC for the Cosmological Constant}

One attractive aspect of the TCC is that it resolves the ``why now'' coincidence problem without anthropic considerations
\cite{Vafa2024,Vafa2025}. Assuming that the Hubble parameter varies slowly and steadily, we obtain
$H \Delta T \lesssim \ln \frac{M_{pl}}{H}$, which implies
\be
\Delta T \lesssim \frac{1}{H} \ln \frac{M_{ {pl}}}{H}.
\ee
This result accounts for the current observational facts regarding the Hubble constant. Furthermore, it implies that when $t\to+\infty$ then $H\to 0^+$, and thus the acceleration of cosmic expansion will stop after a certain epoch.

One motivation for the swampland TCC criterion is the longstanding difficulty in constructing stable de Sitter vacua with a positive cosmological constant \(\Lambda\) from string theory  \cite{Lust2022}. Indeed the TCC extends two earlier major Swampland Conjectures—the Distance Conjecture (SDC)  \cite{Ooguri2006} and the de Sitter Conjecture (dSC)  \cite{ds2018}. The SDC constrains the range of a scalar field by $\Delta \varphi < \frac{M_{pl}}{C} \log \frac{M_{pl}}{\Lambda_{{QG}}}\quad C \sim \mathcal{O}(1)$,
thus setting bounds on the cosmological constant \(\Lambda\) as the vacuum expectation value of certain scalar fields \cite{Scalisi2018}. Moreover, the (refined) de Sitter Conjecture asserts that   stable de Sitter vacua characterized by scalar field potentials $V(\varphi_i)$ are forbidden with a set of scalar fields $\{\varphi_i\}$ contributing to the cosmological constant. Hence, the three conditions i) \(V(\varphi_i)>0\), ii) \( \nabla_{\varphi_i} V = 0\), and iii) \( \min_{i,j}\left(\nabla_{\varphi_i}\nabla_{\varphi_j} V\right)>0\), cannot be satisfied at the same time, and thus only metastable de Sitter vacua that eventually decay are allowed due to de Sitter Conjecture.

It is worth noting that the SDC and dSC are primarily applied to scalar-field realizations of dynamical dark energy. In contrast, the TCC criterion does not presuppose a specific kind of  realization of dark energy, thereby providing a flexible and robust criterion for the validity of a unitary gravitational EFT description. Furthermore, the TCC criterion is believed to be helpful in constructing holographic duality \cite{Bedroya2022,Bedroya2024}.  

In the literature there are many works that  use the TCC   to constrain the early universe features. In the following  we demonstrate that the TCC is also powerful to constrain the behavior of the late-time universe, not only for   dark-energy  models,  but   for modified gravity theories too.  
 

\subsection{TCC applied to Dynamical Dark Energy}\label{sec2.4}

Let us apply the TCC criterion to the expanding history after inflation. In a general dynamical dark energy scenario, with dark energy EoS parameter $w(a)$, the basic evolution is described by Friedmann equation
\be
\frac{H^2}{H_0^2}=\Omega_{de}\exp[-3\int_{1}^{a}\frac{1+w(a^\prime)}{a^\prime}\mathrm{d}a^\prime]+ \Omega_K a^{-2} + \Omega_m a^{-3} + \Omega_r a^{-4},
    \label{fr1}
\ee
where the  dimensionless density parameters $\Omega_i\equiv \rho_{i}/\rho_{crit}$  denote  the density   in terms of the current critical density, and $H_0$ is the current Hubble parameter. A spatially flat universe ($\Omega_K=0$) corresponds to the  current critical density   as $\rho_{crit}=3H_0^2/8\pi G$. In this case,  the TCC criterion can be reformulated as
\begin{equation}
 \Omega_{de}\,\exp\left[-3\int_1^a \frac{1+w(a')}{a'}\,da'\right] < \frac{ M_{pl}^2\,a_{{i}}^2}{H_0^2}\,a^{-2} - \Omega_m\,a^{-3} - \Omega_r\,a^{-4}.
 \label{FullTCC}
 \end{equation}
Similarly, the differentiated version of the TCC criterion $\ddot{a} < 0$ is equivalent to 
\be\begin{aligned}
2\Omega_r\,a^{-4} + \Omega_m\,a^{-3}+ \Omega_{de}\, \exp\left[-3\int_1^a \frac{1+w(a')}{a'}\,da'\right]\,[1+3w(a)] > 0.
\end{aligned}\ee
As we can see, in the very late universe, namely when $t\to +\infty$, where $a$ is sufficiently large or the universe expansion has asymptotically  stopped, we have
\be
w > -\frac{1}{3}.
\ee
Hence, the TCC implies the decelerating fate of the universe\footnote{Oscillations around the averaging behavior are generally possible for   dynamical dark energy scenarios, but they usually denote the entrance of new degrees of freedom and do not affect the overall trend, leading to a smooth effective potential and EoS. Their additional effects will be explored in the future work.}.


Now, according to the differentiated TCC, the condition $\ddot a<0$ must be asymptotically satisfied in the far future. Although our universe has experienced accelerated expansion in the past, it is anticipated that a transition to deceleration will occur at late times. If the dark energy EoS was constant, it would necessarily exceed $w=-1/3$ at all times, which would then preclude any period of acceleration. Consequently, given the current Universe acceleration, the TCC implies that the dark energy EoS must vary with time, and thus dark energy should be dynamical, excluding $\Lambda$CDM and $w$CDM.

We proceed by focusing at late times, and thus we neglect the radiation sector ($\Omega_r\approx  0$). Defining
\be
h(a)=\exp[-3\int_{1}^{a}\frac{1+w(a^\prime)}{a^\prime}\mathrm{d}a^\prime],
\label{hadefinition}
\ee
Eq.\eqref{fr1} can be written as
\be
\frac{H^2}{H_0^2}=[\Omega_{de}h(a) + \Omega_m a^{-3} ].
\ee
According to the TCC  we have
\be
\frac{a}{a_i}\cdot \ell_{pl} \leq \frac{1}{H} \iff a H\leq a_i \ell^{-1}_{pl}.
\label{eq:tcc1}
\ee
Without reference to uncertain early universe physics, the bound on $aH(a)$ requires that sub-Planckian scales at the onset of radiation domination must remain subhorizon in the future. Consequently, one obtains the constraint $aH\leq a_rM_{pl}$, where the subscript $r$ denotes the start of the radiation-dominated era. 

In the following analysis, we focus on dynamical dark energy, assuming that the dark energy contribution was negligible at early cosmological periods, thus preserving the value of $a_r$.
Therefore, the TCC can be expressed as
\be
    \ln \frac{aH(a)}{H_0}=\ln(a)+\frac{1}{2}\ln(\Omega_{de}h(a) + \Omega_m a^{-3} ) \leq \ln(\frac{a_rM_{pl}}{H_0}) \approx 116.4,
    \label{tcconah}
\ee
where we have used that  $a_r = 4.4\times 10^{-11}$ and $H_0=67.4 \mathrm{~km} / \mathrm{s} / \mathrm{Mpc}$, according to Planck 2018\footnote{Since \({a_r M_{{pl}}}/{H_0} \gg 1\), the Hubble Tension does not impact the analysis here.} \cite{Planck:2018vyg}.
 Given the expansion of the Universe ($H>0$), an earlier initial scale factor $a_i$ imposes a more stringent constraint from the TCC. Since the mechanism driving inflation remains unclear and may involve non-perturbative quantum gravity effects, we define the initial time  $a_i$ as the epoch when the $\Lambda$CDM model is presumed to become effective, a value that is estimated by the inverse relation between the scale factor and temperature, namely $a_i=T_0/T_{rh}\approx10^{-29}$ \cite{LiddleLyth2000}, where $T_{rh}$ denotes the reheating temperature. In this context,  Eq.~\eqref{eq:tcc1} becomes
\be
\frac{a}{a_i}\cdot \ell_{pl} = \frac{a}{a_r}\cdot \frac{a_r}{a_i}\cdot  \ell_{pl}  \leq \frac{1}{H},
\ee
and thus Eq.~\eqref{tcconah} becomes
\be
\ln \frac{aH(a)}{H_0} \leq \ln(\frac{a_rM_{pl}}{H_0})-\ln(a_r/a_i) \approx 73.
\label{eq:bound}
\ee
As we observe,  the above relation enables us to impose constraints on the evolution of $H(a)$ and $w(a)$.

\section{Dark-energy Equation of State  Parameterizations  }\label{GR}
\label{sec_3}

In the previous section we applied the TCC criterion on dynamical dark energy, and we extracted condition (\ref{eq:bound}), which can be used to impose constraints on the dark energy equation-of-state parameter  $w(a)$. In this section we proceed to its application to specific dark-energy equation-of-state     parameterizations.

For dynamical dark energy scenarios, the following  three basic constraints   must be satisfied primarily:

\begin{enumerate}
    \item Observational constraint: the EoS should exhibit quintom-B behavior, namely 
    the dark energy EoS $w$ should evolve from \(w < -1\) to \(w > -1\) over time (we do not impose any assumption on whether this occurs at \(a > 1\) or \(a < 1\)).
    \item TCC constraint as \(t\to\infty\): in the infinite future, \(w\) must exceed $-1/3$.  Through numerical computations and quantitative analysis, it can be shown that the TCC violation is unlikely to occur at early times after  inflation, so the requirement for asymptotic behavior at infinity is the concentrated manifestation of  TCC criterion.\footnote{If \(w(a)\) exceeds \(-\frac{1}{3}\) very late, the integral of \(H(t)\) may still exceed \(\ln \frac{M_{\text{pl}}}{H(t_f)}\), thus violating the TCC. This issue will be discussed further in the following part of Section~\ref{GR} and in Appendix \ref{AppenA}.}

    \item Early universe constraint: 
   the early universe must satisfy \( \lim\limits_{a\to 0^+} w(a)<0\) in order to maintain our traditional perception of the expanding history after the inflation, which also allows for a more focused evaluation of the TCC constraints in the late-time universe.
\end{enumerate}

Based on the above requirements, we summarize several common parameterizations  in Table~\ref{table1}, and the observational constraints for these parameterizations have been studied in \cite{Yang:2018qmz,Pan:2019gop,Colgain:2021pmf,Giare:2024gpk,DESI:2025fii}. Since according to the results of  Section \ref{sec2.4},   $\Lambda$CDM and $w$CDM are not favored, our analysis now concentrates on $w_0w_a$CDM, with $w_a\neq0$ in general.

\begin{table}[h]
    \centering
    \resizebox{\textwidth}{!}{
    \begin{tabular}{lcccc}
    \toprule
    \textbf{Model} & \textbf{Functional Form} & \textbf{Allowed by the quintom-B} & \textbf{Allowed by the TCC at $t \to \infty$} & \textbf{Allowed by Early Universe} \\
    \midrule
    CPL \cite{Chevallier:2000qy,Linder:2002et} & $w_0+w_a(1-a)$ & $w_a < 0,w_0+w_a\textless-1$  & $w_a < 0$ & $w_0+w_a < 0$ \\
    
    BA \cite{Barboza:2008rh}  & $w_0+w_a \frac{1-a}{a^2+(1-a)^2}$ & $w_- < -1 < w_+$,   or $w_+<-1<w_0$ & $w_0 > -1/3$ & $w_0+w_a < 0$ \\
    EXP \cite{Dimakis:2016mip,Pan:2019brc} & $w_0-w_a+w_a e^{1-a}$ & $ w_0+w_a(e-1) < -1<w_0-w_a$ & $w_0-w_a > -1/3$  & $w_0+w_a(e-1) < 0$ \\
    LOG \cite{Efstathiou:1999tm} & $w_0-w_a \ln a$ & $w_a < 0$ & $w_a < 0$ & $w_a < 0$ \\
    JBP \cite{Jassal:2005qc} & $w_0+w_a a(1-a)$ & $w_a < 0 , w_0+\frac{w_a}{4}< -1$ & $w_a < 0$ & $w_0 < 0$ \\
    \bottomrule
    \end{tabular}
    }
    \caption{Allowed ranges for various dark-energy equation-of-state  parameterizations.  CPL stands for Chevallier-Polarski-Linder one, BA denotes  the Barboza–Alcaniz one, and JBP the     Jassal-Bagla-Padmanabhan one.}
    \label{table1}
\end{table}

 \subsection{ Chevallier-Polarski-Linder (CPL) parameterization}

\begin{figure*}[t]
\centering
\subfigure[$a_i=10^{-29}$]{
\includegraphics[width=0.51\textwidth]{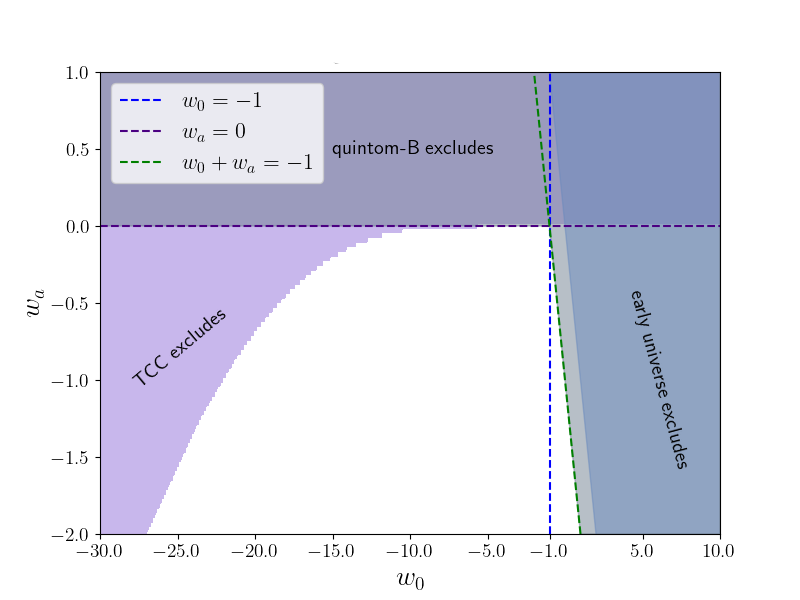}\label{1main}}\subfigure[$a_i=4.4\times10^{-11}$]{
\includegraphics[width=0.51\textwidth]{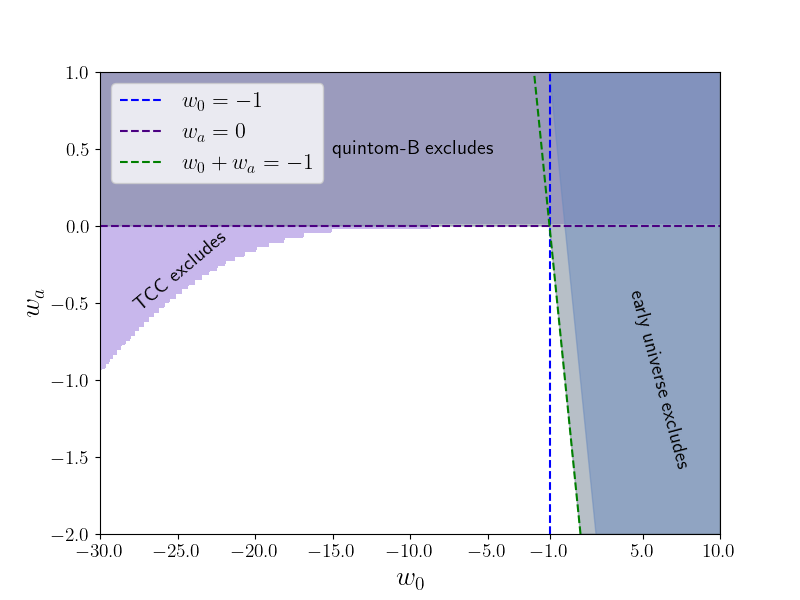}\label{largec}}
\caption{Constraints on CPL dark-energy EoS parameterization (\ref{cplh00}). The purple, gray, and blue region are excluded  by the TCC, the quintom-B scenario, and $w(0)<0$ condition, respectively.}
\label{fig:1}
\end{figure*}

 Let us first examine the Chevallier-Polarski-Linder (CPL) parameterization, which is characterized by the form 
\be
w(a)=w_0+w_a(1-a).
\label{cplh00}
\ee
 In this case, the $\ h(a)$
defined in (\ref{hadefinition})  becomes
\be
h(a)=a^{-3(1+w_0+w_a)}e^{-3w_a(1-a)}.
\label{cplh}
\ee
Substituting Eq.~\eqref{cplh} into Eq.~\eqref{tcconah} and imposing $\Omega_m=0.3,\ \Omega_{de}=0.7$, we extract the constraints on $w_0$ and $w_a$ shown in Fig.~\ref{fig:1}. 

\begin{figure}[htbp]
    \centering    \includegraphics[width=0.85\textwidth]{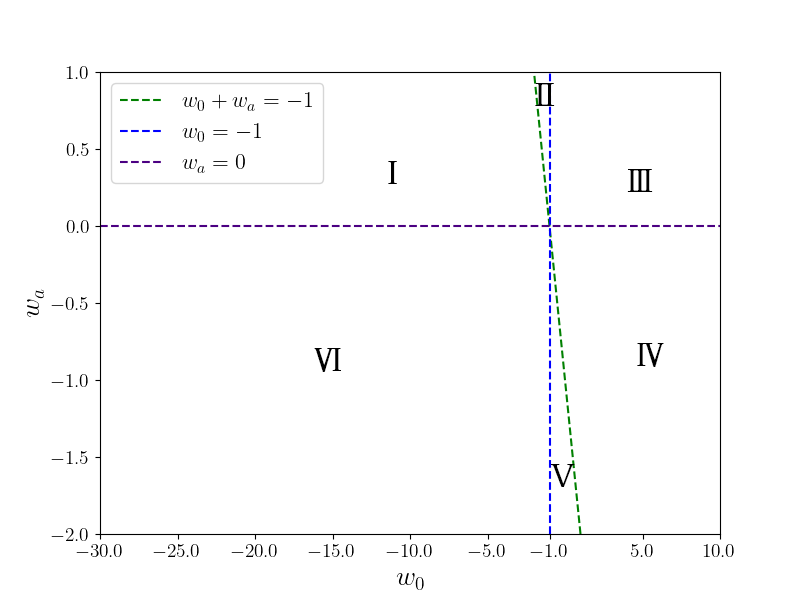}
	\caption{
     We divide the  $w_0-w_a$ plane into six parts, by the lines of $w_0 = -1$, $w_0 +w_a =-1$
 and $w_a = 0$. Part \uppercase\expandafter{\romannumeral1} corresponds to the phantom scenario; Part \uppercase\expandafter{\romannumeral2} and \uppercase\expandafter{\romannumeral5} to quintom-A and   quintom-B scenarios with an EoS that has already crossed $-1$; Part \uppercase\expandafter{\romannumeral3} and \uppercase\expandafter{\romannumeral6} to   quintom-A and   quintom-B scenarios with a future crossing of $-1$; and Part \uppercase\expandafter{\romannumeral4} to the quintessence scenario.}
    \label{fig:region}
\end{figure}

Based on the basic constraints described above, we deduce that the CPL parameterization is required to satisfy $w_0+w_a\textless-1$ and $w_a\textless0$. Meanwhile, the TCC forbids $w$ from crossing $-1$ too late, resulting to an exclusion region for $w_a<0$, as illustrated in Fig.~\ref{fig:1}. In other words, although the TCC requires $w$ to exceed $-1/3$ as $t\to\infty$, our numerical analysis reveals that parameter regions where $w$ exceeds $-1/3$ too late should be excluded too. Additionally, Fig.~\ref{largec} and Fig.~\ref{1main} indicate  that the value of $a_i$ has little impact on the region where the cosmological constraints from the data are concentrated. Our analysis shows that the DESI results  \cite{DESI:2024mwx,DESI:2025zgx} lie within the region allowed by the TCC. This suggests that although the DESI indication of dynamical dark energy favoring a quintom-B scenario may be astonishing, it still remains compatible with the TCC. Nevertheless, the TCC alone favors scenarios in which dark energy weakens over time, without making a definitive statement regarding the quintom-B behavior.

Proceeding forward, in Fig.~\ref{fig:region} we present the parameter space corresponding to various dynamical dark energy behaviors under the CPL parametrization, using the parameter choices of Fig.~\ref{fig:1}. A comparison of these two figures confirms our previous analysis: within the CPL framework, the TCC favors either the quintessence scenario or quintom-B models that do not cross $-1$ too late.

We mention that the CPL parameterization coincides with first-order Taylor expansion of $w(a)$ around the present epoch $a =1$, which means it is inadequate at $a\gg 1 $. Therefore, it would be better to 
consider  parameterizations that are convergent at high $a$, such as the Barboza-Alcaniz parametrization.

 \subsection{ Barboza-Alcaniz (BA)  parameterization}
 
The  Barboza-Alcaniz (BA) parametrization is characterized by the form
\begin{align}
    w(a)=w_0+w_a\frac{(1-a)}{a^2+(1-a)^2},
    \label{eq:BA}
\end{align}
which inserted into (\ref{hadefinition})  yields:
\be
h(a)=a^{-3(1+w_0+w_a)}(1-2a+2a^2)^{\frac{3}{2}w_a}.
\ee
As we can see, the BA EoS   Eq.~\eqref{eq:BA} exhibits extrema at $a^*_{\pm} = 1\pm \sqrt{2}/2$, which correspond to $w_- = w(a^*_-) = w_0+ \frac{1+\sqrt{2}}{2}w_a\approx w_0 + 1.21w_a$ and $w_+ = w(a^*_+) = w_0+\frac{1-\sqrt{2}}{2}w_a\approx w_0 - 0.21w_a$. To achieve a quintom-B behavior, there are two possibilities: i) $w_{-}<-1<w_{+}$, with $w_a<0$; ii) $w_{+}<-1<w_0$, with $w_a>0$. Combining with other two requirements, the regime $w_a>0$ is excluded. For $w_a < 0$, $w_+$ is a maximum, and $w_-$ is a minimum. In order to avoid uncontrolled accelerated expansion, the asymptotic value must satisfy $w(a\to \infty)=w_0\textgreater -\frac{1}{3}$.

\begin{figure*}[t]
\centering
\subfigure{
    \includegraphics[width=0.5\textwidth]{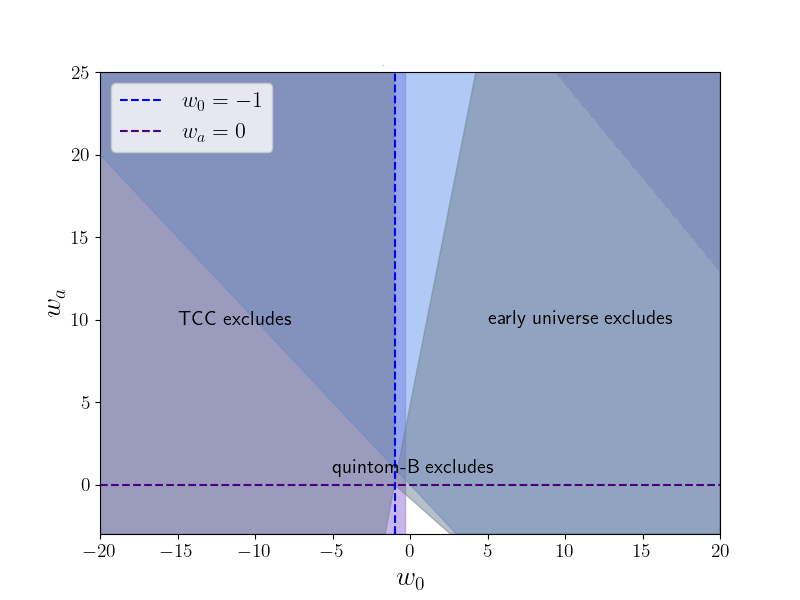}\label{BA1}}\subfigure{
\includegraphics[width=0.5\textwidth]{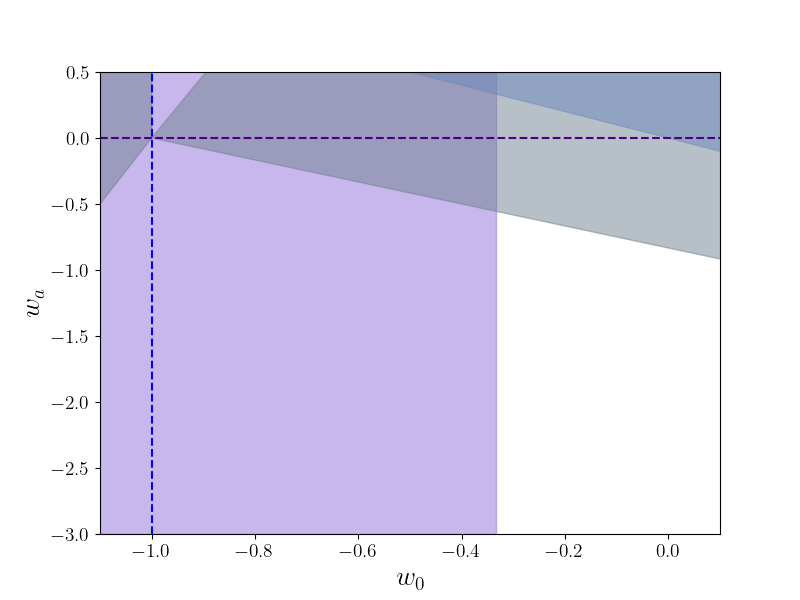}\label{BA2}}
\caption{Left panel: Constraints on the BA  dark-energy EoS parameterization (\ref{eq:BA}) by setting $a_i=10^{-29}$. The purple, gray, and blue regions are excluded  by the TCC, the quintom-B scenario, and    $w(0)<0$ condition, respectively. Right panel: enlarged region where the cosmological constraints from the data are zoomed in. The legend is the same as the left panel.}
\label{fig:BA}
\end{figure*}

In  Fig.~\ref{fig:BA} we present the corresponding constraints. As we can see, 
  our analysis reveals that the upper bound of $w_0$ approaches $-1/3$. Moreover, when $w_a\textgreater 0$, the minimum value of $w$ cannot be too low, imposing an upper bound on $w_a$, since excessively extreme or prolonged accelerated expansion has also been precluded.  Notably, a substantial region of the parameter space corresponding to the quintom-B scenario is excluded too by the TCC. It seems that there is a strong constraint on observations that excludes $w(a=1)=w_0<-1/3$. However, this is primarily due to the BA form assumption that $w(a=1)=w(a\to\infty)=w_0$, which lacks physical motivation.

\subsection{EXP, LOG, and JBP parameterization}

There are  other parameterizations that have been tested in previous data analyses too \cite{DESI:2025fii}, such as the exponential (EXP) parameterization $w(a)=w_0-w_a+w_a e^{1-a}$, the logarithmic (LOG) parameterization $w(a)=w_0-w_a \ln a$, and the Jassal–Bagla–Padmanabhan (JBP) parameterization $w(a)=w_0+w_a a(1-a)$.

Plugging these three parameterizations into Eq.~(\ref{hadefinition}), we obtain
\be
\begin{aligned}
    {\rm EXP}: h(a)&=a^{-3\left(1+w_0-w_a\right)}\exp\left[3w_a e\left(E_1(a)-E_1(1)\right)\right] \\
{\rm LOG}: h(a)&=a^{-3\left(1+w_0\right)}\exp\left[\frac{3w_a}{2} \left(\ln a\right)^2\right] \\
{\rm JBP}: h(a)&=a^{-3\left(1+w_0\right)}\exp\left[\frac{3w_a}{2} \left(a-1\right)^2\right] \\
\end{aligned},
\label{EXPh}
\ee
where $E_1(x)\equiv \Gamma(0, x)\equiv \int_x^{+\infty}\frac{e^{-x}}{x}dx$ is the exponential integral function. For the EXP, LOG, and JBP parameterizations, the choice of $(w_0,w_a)$ should avoid that dark energy density \(\Omega_{de}\, h(a)\) diverges super-polynomially as \(a\to 0^+\) or \(a\to +\infty\). According to the Friedmann equation~(\ref{fr1}), such divergence naturally contradicts early universe observations or fails to satisfy the TCC criterion, due to an uncontrolled expansion rate.  The conclusion on the parameter bounds is shown in Table~\ref{table1}.

In summary, although these parameterizations provide good approximations at low redshift, their divergence at high redshifts, or high blueshift indicates that they are, in essence, effective parameterizations valid only under low-redshift conditions rather than fundamental descriptions of dark energy behavior.

From the above analysis, we can conclude that the region allowed by the TCC significantly overlaps with that of the quintom-B scenario, which coincides with the parameter space favored by current observations. Although these observational results may appear unexpected, they do not place the universe in the swampland. Similarly, a quintessence model with $w\textless0$ in the early universe — such as thawing dark energy \cite{Caldwell:2005tm,Wolf:2023uno} —   lies within the landscape of TCC, too. 

We close this section by mentioning that different parameterizations yield comparable confidence levels in observations \cite{DESI:2025fii}, exhibiting similar quasi-linear behaviors at low redshift $w(a) \approx\left(w_0^{\mathrm{eff}}+w_a^{\mathrm{eff}}\right)-w_a^{\mathrm{eff}} a$, and predicting quintom-B behavior at approximately the same epoch. Current experimental precision, however, remains insufficient to distinguish among these parameterizations, largely due to the scarcity of high-redshift data. The TCC criterion will impose significantly stronger constraints on dynamical dark energy scenarios as high-redshift observations accumulate and more refined parameterizations are developed.

\section{Modified Gravity}\label{MG}
\label{sec_4}

As we mentioned in the Introduction, modified gravity is the second main avenue that one can follow in order to describe an  effective dark-energy sector. In this section we are interested in applying the TCC criterion for some widely-used models of modified gravity from the torsional and non-metricity classes.

\subsection{$f(T)$ and  $f(Q)$ gravities}

As is known, one can adopt a geometrical approach to deviate from General Relativity and thus modify the $\Lambda$CDM
 paradigm.  
 Geometric effects from modification of Einstein gravity may lead to an apparent violation of the NEC and induce quintom behavior, while the theory itself remains well defined. Furthermore, most modified gravity models are required to recover the predictions of general relativity at small scales, ensuring that the validity of the singularity theorems is not compromised.

Two widely investigated classes of modified gravity theories are the $f(T)$ gravity \cite{Cai2015,Bahamonde:2021gfp} and the $f(Q)$ gravity \cite{BeltranJimenez:2019tme}. 
 These theories are characterized by 
 the gravitational action  
\be
S=\frac{1}{16 \pi G}\int d^4 x \sqrt{-g}\left[  f(X)\right],
\ee
where $X$ can be either the torsion scalar $T$ or the non-metricity scalar $Q$,   defined respectively as
\begin{align}
T &= \frac{1}{4} T^\rho{ }_{\mu \nu} T_\rho{ }^{\mu \nu} + \frac{1}{2} T^\rho{ }_{\mu \nu} T^{\nu \mu}{ }_\rho - T^\rho{ }_{\mu \rho} T_\nu{ }^{\nu \mu}, 
\end{align}
\begin{align}
Q &=  \frac{1}{4} Q_{\alpha \mu \nu} Q^{\alpha \mu \nu} - \frac{1}{2} Q_{\alpha \mu \nu} Q^{\mu \alpha \nu} 
 - \frac{1}{4} \left(g^{\mu \nu} Q_{\alpha \mu \nu}\right)\left(g^{\alpha \beta} g^{\rho \sigma} Q_{\beta \rho \sigma}\right) \nonumber\\
 & \ \   \
 + \frac{1}{2} \left(g^{\mu \nu} Q_{\alpha \mu \nu}\right)\left(g^{\alpha \beta} g^{\rho \sigma} Q_{\rho \beta \sigma}\right),
\end{align}
 
where $T^\alpha{ }_{\mu \nu} = \Gamma^\alpha{ }_{\nu \mu} - \Gamma^\alpha{ }_{\mu \nu}$ is the torsion tensor, and $Q_{\alpha \mu \nu} = \nabla_\alpha g_{\mu \nu}$ is the non-metricity tensor.  When $f(X) = X$, both $f(T)$ and $f(Q)$ gravity theories reduce to General Relativity, revealing the  ``gravity trinity'' \cite{Heisenberg:2018vsk,BeltranJimenez:2019esp}, and one advantage of $f(T), f(Q)$ gravities compared with the $f(R)$ gravity  is that in the 
Friedman-Lema\^{i}tre-Robertson-Walker (FLRW) geometry they do not involve derivatives higher than second order in the equation of motion.

Since the equivalence between $f(T)$ and $f(Q)$ theories (with the coincident gauge, namely the simplest connection that inherits the symmetries of the
background spacetime) in the 
FLRW geometry has been discussed in detail in the literature  \cite{Wu:2024vcr,Basilakos:2025olm}, we mention here that  in the following   we focus on the $f(T)$ gravity theory however the results hold for $f(Q)$ gravity as well, under the substitution $T\rightarrow Q$.

Let us  make the split   $f(T)=T + F(T)$. In this case  
the modified Friedmann equations  become 
\begin{align}
H^2 &= \frac{8 \pi G }{3}\rho_{m} -\frac{F}{6}-2H^2 F_T \,,
\label{eq:frft}
 \\ 
\frac{d H^2}{d\ln a}  &=
\frac{16 \pi GP_{m} 
+ 6H^2+F+12H^2F_T}{24H^2F_{TT}-2-2F_T}\,, 
\end{align}
  where the torsion scalar is   given by $T = -6H^2$, and where 
  $F_T \equiv dF(T)/dT$  and $F_{TT} \equiv d^2F(T)/dT^2$. Moreover, 
 $\rho_{m}$ and $P_{m}$ are
the energy density and pressure of  the matter  perfect fluid, respectively. 
Hence, we can re-write these equations in the standard form,  namely
\begin{align}
H^2 &= \frac{8 \pi G }{3} \left(\rho_{m}+\rho_{{de}} 
\right)\,, 
\\
\frac{d H^2}{d\ln a}  
&= -8 \pi G \left(\rho_{m} + P_{m} + 
\rho_{{de}} + P_{{de}} \right)\,,
\label{eq:frgr} 
\end{align}
by introducing  an effective dark energy  sector with energy density and pressure   given by
\begin{align}
\rho_{{de}} =& \frac{1}{16 \pi G}\left(-F+2TF_{T}\right)\,,
\\ 
P_{{de}} =&
\frac{1}{16 \pi G} \frac{F-TF_T+2T^2F_{TT}}{1+F_T+2TF_{TT}}\,.
\label{rhop} 
\end{align}
Consequently, the effective dark-energy  equation-of-state  parameter is written as
\be
w_{{de}} \equiv \frac{P_{{de}}}{\rho_{{de}}} 
=\frac{F/T-F_T+2TF_{TT}}{
\left(1+F_T+2TF_{TT}\right)\left(F/T-2F_T\right)}\,.
\label{eq:wft} 
\ee

\subsection{Specific  $f(T)$ gravity models}\label{ft1}

We proceed by considering   specific $f(T)$ gravity  models. 
We begin by considering the exponential $f (T)$ theory introduced in  \cite{Linder_2010}, which is given by
\be
F(T)= \alpha T \left(1-e^{pT_0/T}\right),
\label{eq:F} 
\ee
with
\be
\alpha = -\frac{1-\Omega_{m}}{1-\left(1-2p\right)e^p} .
\label{eq:alpha} 
\ee
Here, $p$ is a constant (with $p=0$ corresponding to the $\Lambda$CDM model) and $T_0=T(a=1)$ denotes the current torsion scalar \cite{Nesseris:2013jea}.  
By setting $E=\frac{T}{T_0}=\frac{H^2}{H_0^2}$ and inserting into  \eqref{eq:frft}, we obtain
\begin{align} 
   E &= \Omega_{m} a^{-3} + \frac{F}{T_0} -2F_T E.
    \label{eq:x}
\end{align}
Note that in the limit $a\gg 1$, $E$ only involves the single dimensionless parameter $p$.

Let us now apply the TCC criterion. This imposes the bound 
\be
    \ln a+\frac{1}{2}\ln E \leq \ln(\frac{a_iM_{pl}}{H_0}) ,
\ee
which then provides the constraints on the 
  parameter $p$. As we observe, for this model  the universe resides either in a phantom phase ($p \textgreater 0$) or in a quintessence phase ($p \textless 0$), with no phantom-divide crossing realization. In particular, when $p\textgreater0$  the value of dark energy EoS remains strictly below $-1$, characterizing a phantom regime that violates the TCC. In contrast, for $p\textless0$, only sufficiently small values of $p$ allow for a decelerating expansion. This behavior is illustrated in Fig. 1 of  \cite{Bamba_2011}.
 
Since under the exponential model  the universe remains in either the phantom or the quintessence phase,  we proceed to the investigation of the cosmological evolution in a combined $f(T)$ theory, with both logarithmic and exponential terms, which allows    the quintom behavior.
The explicit form of theory is given by  \cite{Bamba_2011}
\be
F(T)= \gamma \left[ T_0 \left(\frac{uT_0}{T}\right)^{-1/2} \ln 
\left(\frac{uT_0}{T}\right) 
-T \left(1-e^{uT_0/T}\right) 
\right] ,
\label{eq:ftcom} 
\ee
with
\be
\gamma \equiv \frac{1-\Omega_{m}}{2u^{-1/2}
+\left[1-\left(1-2u\right)e^u \right]}\,,
\label{eq:ftqa}
\ee
where $u$ is a constant. Here, we   restrict to $u >0$ to simplify the analysis
when combining the two terms. In this case, the first Friedmann equation leads to the same expression as Eq.~\eqref{eq:x}, and thus when $a$ is large enough it is independent of $u$.

Notably, independently of the parameter $u$, the model \eqref{eq:ftcom} characterizes a quintom-A model (see Fig. 7 of  \cite{Bamba_2011}), which implies that the universe undergoes permanent acceleration, ultimately violating the TCC when the scale factor becomes sufficiently large.

 \subsection{TCC violation and modified gravity}

In the previous subsection we tested two well-used $f(T)$ (and thus $f(Q)$) modified gravity models and we found that they    tend to     violate the constraints imposed by the TCC. This reveals that modified gravity theories may be subject to equally strong restrictions with the dark-energy parameterizations in the framework of General Relativity.

In fact, this behavior can be reasonably explained. According to the Lovelock theorem \cite{Lovelock1971}, a gravity theory in   $D=4$ dimensions, depending only on the metric \(g_{\mu \nu}\) and its derivatives up to second order, admits no non-trivial modification of General Relativity other than the addition of a cosmological constant $\Lambda$ with EoS $w=-1$. When combined with the cosmological principle and the observed late-time acceleration of the universe, this naturally leads to the \(\Lambda\)CDM scenario. Therefore, unless the cosmological constant is extended as a form of dynamical dark energy, or as an effective term arising from gravitational modification, it would be difficult to satisfy the TCC constraints in the distant future.

In the case of modified gravity, although the geometry may not be pseudo-Riemannian, according to gravitational trinity \cite{Heisenberg:2018vsk,BeltranJimenez:2019esp}, the theory will recover Einstein gravity in the infrared (IR) limit since the modifications to Einstein gravity always arise at higher orders by simple dimensional analysis at the classical level. Consequently, the modifications will be  suppressed at the cosmological scales and thus they cannot provide significant corrections at late epochs compared with Einstein gravity, such as those demonstrated in Fig. 6 of \cite{Cai2015} for exponential $f(T)$ gravity, and also supported by linear and second order perturbative calculations \cite{Hu2023, Hu2023xcf}. This implies that IR modifications to Einstein gravity at very large scales should be considered seriously, and thus models such as nonlocal gravity \cite{Maggiore2014}, non-relativistic gravity \cite{Cai2009}  and Hořava–Lifshitz gravity \cite{Horava2009} are worthwhile revisited, as well as other possibilities including interacting dark energy \cite{wang2016} or the unification of the dark sector \cite{Bilic2001,Bento2002,Manous2017,Vafa2024fpx}.

\section{Conclusion}\label{conclusion}
\label{sec_5}

In this study we examined the dynamics of dark energy by analyzing its equation-of-state parameter. We focused on scenarios where the dark energy weakens during its evolution, as indicated by the latest  DESI DR2  data, and we examined them within the framework of quantum gravity effective field theory constraints. In particular, it is known that  according to the Trans-Planckian Censorship Conjecture, any trans-Planckian quantum fluctuation should not 
stretch beyond the
Hubble radius during cosmic expansion, since this would imply that unknown trans-Planckian physics would affect the low-energy behavior of our theories, and thus  our effective field theories  would not be “effective field theories”  with trustworthy predictions. Hence, according to TCC, the Universe expansion cannot be accelerating for ever and hence dark energy should be dynamical in a suitable way that forbids this possibility.

Under the light of TCC we examined various established EoS parameterizations, specifically the CPL, BA, EXP, LOG, and JBP forms, and we extracted the allowed parameter space under the combined stipulations of the TCC and quintom criterion. We elucidated the constrained parameter space and we analyzed the compatibility between observations and the TCC.

Additionally, we investigated  the feasibility of specific models of $f(T)$ and $f(Q)$ modified gravity theories concerning the TCC criterion. As we showed, the  two   well-known models we examined violate the constraints imposed by the TCC. Hence, 
TCC enforces equally rigorous constraints on modified gravity theories, effectively excluding a broad class of modified gravity theories that asymptotically converge to $\Lambda$CDM in the asymptotically future.

Notably, the energy scale of the late universe is significantly lower than the Planck scale, rendering the TCC constraints in this regime more robust than those applicable to the early universe.
As a result, models that predict sustained accelerated expansion into the distant future are effectively excluded.
Additionally, since quintom behavior violates the NEC within the \(\Lambda\)CDM framework, applying these constraints simultaneously significantly narrows the range of viable models. If further observations confirm the quintom-B behavior, this integrated approach will provide clear guidance for the development of new physics.

\acknowledgments
We are grateful for Qinxun Li, Xin Ren, Shao-Jiang Wang, Zhi-Zhen Wang and Yuhang Yang  for insightful discussions. This work was supported in part by the National Key R\&D Program of China (2021YFC2203100, 2024YFC2207500), by the National Natural Science Foundation of China (12433002, 12261131497, 92476203), by CAS young interdisciplinary innovation team (JCTD-2022-20), by 111 Project (B23042), by USTC Fellowship for International Cooperation, and by USTC Research Funds of the Double First-Class Initiative. ENS acknowledges the contribution of the LISA CosWG and the COST Actions  and  of COST Actions CA21136 ``Addressing observational tensions in cosmology with systematics and fundamental physics (CosmoVerse)'',  CA21106 ``COSMIC WISPers in the Dark Universe: Theory, astrophysics and experiments (CosmicWISPers)'', and CA23130 ``Bridging high and low energies in search of quantum gravity (BridgeQG)''. JW acknowledges YMSC and TSIMF for hospitality and a stimulating intellectual environment during the workshop ``The Lotus and Swampland''.

\appendix

\section{Analysis of the  Trans-Planckian Censorship Conjecture for Dark Energy Dominating Epochs}
\label{AppenA}
 
According to Eq.~(\ref{FullTCC}), the complete TCC constraint for the \(w_0w_a\)CDM parameterizations can be formulated as  
\be
\mathscr{F}_{\text{Para}}\left(a, w_0, w_a\right) +\Omega_m a^{-1}+\Omega_r a^{-2}< C_0, \quad a\in (a_i, +\infty),
\ee
where “Para” denotes a specific parameterization and we have the generic form
$
\mathscr{F}_{{\rm Para}}\left(a, w_0, w_a\right) = a^2\,\Omega_{de}\, h(a)
$. Here \(C_0 = \frac{M_{ {pl}}^2\,a_{i}^2}{H_0^2}\).

Let \(a_{ max}\) denote the value of \(a\) corresponding to the maximum of \(\mathscr{F}_{{\rm Para}}(a, w_0, w_a)\).  Suppose the possible violation of TCC happens in the late universe when dynamical dark energy dominates, we have:

\begin{enumerate}
    \item[(i)] When \(a_{ max} \le a_i\), the TCC is automatically satisfied.
    \item[(ii)] When \(a_{ max} > a_i\), the TCC for the dark energy-dominated epoch is equivalent to 
    \be
    \mathscr{F}_{\rm Para}(a_{ max}, w_0, w_a) <  {C_0}.
    \ee
\end{enumerate}

 Based on parameter spaces allowed in Table~\ref{table1}, the positions of the maxima for the \(w_0w_a\) parameterizations can be calculated as follows.

\subsection*{CPL Parameterization}

For the CPL parameterization, the maximum occurs at 
\be
a_{ max} = \frac{1+3w_0+3w_a}{3w_a} > 0.
\ee

\subsection*{BA Parameterization}

For the BA parameterization, the maximum occurs at 
\be
a_{ max} = \frac{2(1+3w_0)+3w_a+\sqrt{-4(1+3w_0)^2-12(1+3w_0)w_a+9w_a^2}}{4(1+3w_0)} > 0.
\ee

\subsection*{JBP Parameterization}

For the JBP parameterization, the maximum occurs at 
\be
a_{ max} = \frac{1+\sqrt{1+\frac{4(1+3w_0)}{3w_a}}}{2} > 0.
\ee

\subsection*{LOG Parameterization}

For the LOG parameterization, the maximum occurs at 
\be
a_{ max} = \exp\left(\frac{1+3w_0}{3w_a}\right) > 0.
\ee
Notably, the maximum for the LOG parameterization assumes a particularly simple form, namely $\mathscr{F}_{\rm {LOG}}\left(a_{ max}, w_0, w_a\right) = \exp\left[-\frac{(1+3w_0)^2}{6w_a}\right]$.

\subsection*{EXP Parameterization}

For the EXP parameterization, the maximum occurs at
\be
a_{ max} = \ln\left(\frac{w_a e}{w_a-w_0-1/3} \right) > 0.
\ee

\bibliographystyle{JHEP}
\bibliography{TCC}

\providecommand{\href}[2]{#2}\begingroup\raggedright\begin{thebibliography}{10}

\bibitem{SupernovaSearchTeam:1998fmf}
{\scshape Supernova Search Team} collaboration, \emph{{Observational evidence from supernovae for an accelerating universe and a cosmological constant}}, \href{https://doi.org/10.1086/300499}{\emph{Astron. J.} {\bfseries 116} (1998) 1009} [\href{https://arxiv.org/abs/astro-ph/9805201}{{\ttfamily astro-ph/9805201}}].

\bibitem{SupernovaCosmologyProject:1998vns}
{\scshape Supernova Cosmology Project} collaboration, \emph{{Measurements of $\Omega$ and $\Lambda$ from 42 High Redshift Supernovae}}, \href{https://doi.org/10.1086/307221}{\emph{Astrophys. J.} {\bfseries 517} (1999) 565} [\href{https://arxiv.org/abs/astro-ph/9812133}{{\ttfamily astro-ph/9812133}}].

\bibitem{DESI:2024mwx}
{\scshape DESI} collaboration, \emph{{DESI 2024 VI: cosmological constraints from the measurements of baryon acoustic oscillations}}, \href{https://doi.org/10.1088/1475-7516/2025/02/021}{\emph{JCAP} {\bfseries 02} (2025) 021} [\href{https://arxiv.org/abs/2404.03002}{{\ttfamily 2404.03002}}].

\bibitem{DESI:2025zgx}
{\scshape DESI} collaboration, \emph{{DESI DR2 Results II: Measurements of Baryon Acoustic Oscillations and Cosmological Constraints}},  \href{https://arxiv.org/abs/2503.14738}{{\ttfamily 2503.14738}}.

\bibitem{DESI:2025fii}
{\scshape DESI} collaboration, \emph{{Extended Dark Energy analysis using DESI DR2 BAO measurements}},  \href{https://arxiv.org/abs/2503.14743}{{\ttfamily 2503.14743}}.

\bibitem{DESI:2025ejh}
{\scshape DESI} collaboration, \emph{{Constraints on Neutrino Physics from DESI DR2 BAO and DR1 Full Shape}},  \href{https://arxiv.org/abs/2503.14744}{{\ttfamily 2503.14744}}.

\bibitem{Gongbo2025}
G.~Gu et~al., \emph{{Dynamical Dark Energy in light of the DESI DR2 Baryonic Acoustic Oscillations Measurements}},  \href{https://arxiv.org/abs/2504.06118}{{\ttfamily 2504.06118}}.

\bibitem{Yin2025}
J.~Lee, K.~Murai, F.~Takahashi and W.~Yin, \emph{{Isotropic cosmic birefringence from string axion domain walls without cosmic strings, and DESI results}},  \href{https://arxiv.org/abs/2503.18417}{{\ttfamily 2503.18417}}.

\bibitem{Feng:2004ad}
B.~Feng, X.-L.~Wang and X.-M.~Zhang, \emph{{Dark energy constraints from the cosmic age and supernova}}, \href{https://doi.org/10.1016/j.physletb.2004.12.071}{\emph{Phys. Lett. B} {\bfseries 607} (2005) 35} [\href{https://arxiv.org/abs/astro-ph/0404224}{{\ttfamily astro-ph/0404224}}].

\bibitem{Huterer:2004ch}
D.~Huterer and A.~Cooray, \emph{{Uncorrelated estimates of dark energy evolution}}, \href{https://doi.org/10.1103/PhysRevD.71.023506}{\emph{Phys. Rev. D} {\bfseries 71} (2005) 023506} [\href{https://arxiv.org/abs/astro-ph/0404062}{{\ttfamily astro-ph/0404062}}].

\bibitem{Cai:2009zp}
Y.-F.~Cai, E.N.~Saridakis, M.R.~Setare and J.-Q.~Xia, \emph{{Quintom Cosmology: Theoretical implications and observations}}, \href{https://doi.org/10.1016/j.physrep.2010.04.001}{\emph{Phys. Rept.} {\bfseries 493} (2010) 1} [\href{https://arxiv.org/abs/0909.2776}{{\ttfamily 0909.2776}}].

\bibitem{Wetterich:1994bg}
C.~Wetterich, \emph{{The Cosmon model for an asymptotically vanishing time dependent cosmological 'constant'}}, {\emph{Astron. Astrophys.} {\bfseries 301} (1995) 321} [\href{https://arxiv.org/abs/hep-th/9408025}{{\ttfamily hep-th/9408025}}].

\bibitem{Amendola:1999er}
L.~Amendola, \emph{{Coupled quintessence}}, \href{https://doi.org/10.1103/PhysRevD.62.043511}{\emph{Phys. Rev. D} {\bfseries 62} (2000) 043511} [\href{https://arxiv.org/abs/astro-ph/9908023}{{\ttfamily astro-ph/9908023}}].

\bibitem{Saridakis:2012jy}
E.N.~Saridakis, \emph{{Phantom crossing and quintessence limit in extended nonlinear massive gravity}}, \href{https://doi.org/10.1088/0264-9381/30/7/075003}{\emph{Class. Quant. Grav.} {\bfseries 30} (2013) 075003} [\href{https://arxiv.org/abs/1207.1800}{{\ttfamily 1207.1800}}].

\bibitem{CANTATA:2021asi}
{\scshape CANTATA} collaboration, E.N.~Saridakis et~al., \emph{{Modified Gravity and Cosmology. An Update by the CANTATA Network}}, Springer (2021), \href{https://doi.org/10.1007/978-3-030-83715-0}{10.1007/978-3-030-83715-0}, [\href{https://arxiv.org/abs/2105.12582}{{\ttfamily 2105.12582}}].

\bibitem{Yang:2024kdo}
Y.~Yang, X.~Ren, Q.~Wang, Z.~Lu, D.~Zhang, Y.-F.~Cai et~al., \emph{{Quintom cosmology and modified gravity after DESI 2024}}, \href{https://doi.org/10.1016/j.scib.2024.07.029}{\emph{Sci. Bull.} {\bfseries 69} (2024) 2698} [\href{https://arxiv.org/abs/2404.19437}{{\ttfamily 2404.19437}}].

\bibitem{Yang:2025kgc}
Y.~Yang, Q.~Wang, C.~Li, P.~Yuan, X.~Ren, E.N.~Saridakis et~al., \emph{{Gaussian-process reconstructions and model building of quintom dark energy from latest cosmological observations}},  \href{https://arxiv.org/abs/2501.18336}{{\ttfamily 2501.18336}}.

\bibitem{Yang:2025mws}
Y.~Yang, Q.~Wang, X.~Ren, E.N.~Saridakis and Y.-F.~Cai, \emph{{Modified gravity realizations of quintom dark energy after DESI DR2}},  \href{https://arxiv.org/abs/2504.06784}{{\ttfamily 2504.06784}}.

\bibitem{Giare:2024smz}
W.~Giar\`e, M.A.~Sabogal, R.C.~Nunes and E.~Di~Valentino, \emph{{Interacting Dark Energy after DESI Baryon Acoustic Oscillation Measurements}}, \href{https://doi.org/10.1103/PhysRevLett.133.251003}{\emph{Phys. Rev. Lett.} {\bfseries 133} (2024) 251003} [\href{https://arxiv.org/abs/2404.15232}{{\ttfamily 2404.15232}}].

\bibitem{Li:2024qso}
T.-N.~Li, P.-J.~Wu, G.-H.~Du, S.-J.~Jin, H.-L.~Li, J.-F.~Zhang et~al., \emph{{Constraints on Interacting Dark Energy Models from the DESI Baryon Acoustic Oscillation and DES Supernovae Data}}, \href{https://doi.org/10.3847/1538-4357/ad87f0}{\emph{Astrophys. J.} {\bfseries 976} (2024) 1} [\href{https://arxiv.org/abs/2407.14934}{{\ttfamily 2407.14934}}].

\bibitem{Zhai:2025hfi}
Y.~Zhai, M.~de~Cesare, C.~van~de Bruck, E.~Di~Valentino and E.~Wilson-Ewing, \emph{{A low-redshift preference for an interacting dark energy model}},  \href{https://arxiv.org/abs/2503.15659}{{\ttfamily 2503.15659}}.

\bibitem{Pan:2025qwy}
S.~Pan, S.~Paul, E.N.~Saridakis and W.~Yang, \emph{{Interacting dark energy after DESI DR2: a challenge for $\Lambda$CDM paradigm?}},  \href{https://arxiv.org/abs/2504.00994}{{\ttfamily 2504.00994}}.

\bibitem{Ye:2024ywg}
G.~Ye, M.~Martinelli, B.~Hu and A.~Silvestri, \emph{{Non-minimally coupled gravity as a physically viable fit to DESI 2024 BAO}},  \href{https://arxiv.org/abs/2407.15832}{{\ttfamily 2407.15832}}.

\bibitem{Pan:2025psn}
J.~Pan and G.~Ye, \emph{{Non-minimally coupled gravity constraints from DESI DR2 data}},  \href{https://arxiv.org/abs/2503.19898}{{\ttfamily 2503.19898}}.

\bibitem{Cai:2019igo}
R.-G.~Cai and S.-J.~Wang, \emph{{Mass bound for primordial black hole from trans-Planckian censorship conjecture}}, \href{https://doi.org/10.1103/PhysRevD.101.043508}{\emph{Phys. Rev. D} {\bfseries 101} (2020) 043508} [\href{https://arxiv.org/abs/1910.07981}{{\ttfamily 1910.07981}}].

\bibitem{Payeur:2024kyy}
G.~Payeur, E.~McDonough and R.~Brandenberger, \emph{{Swampland conjectures constraints on dark energy from a highly curved field space}}, \href{https://doi.org/10.1103/PhysRevD.110.106011}{\emph{Phys. Rev. D} {\bfseries 110} (2024) 106011} [\href{https://arxiv.org/abs/2405.05304}{{\ttfamily 2405.05304}}].

\bibitem{Arjona:2024dsr}
R.~Arjona and S.~Nesseris, \emph{{A swampland conjecture DESIder\'atum?}},  \href{https://arxiv.org/abs/2409.14990}{{\ttfamily 2409.14990}}.

\bibitem{Bhattacharya:2024kxp}
S.~Bhattacharya, G.~Borghetto, A.~Malhotra, S.~Parameswaran, G.~Tasinato and I.~Zavala, \emph{{Cosmological tests of quintessence in quantum gravity}},  \href{https://arxiv.org/abs/2410.21243}{{\ttfamily 2410.21243}}.

\bibitem{Brandenberger:2025hof}
R.~Brandenberger, \emph{{Why the DESI Results Should Not Be A Surprise}},  \href{https://arxiv.org/abs/2503.17659}{{\ttfamily 2503.17659}}.

\bibitem{Anchordoqui:2025fgz}
L.A.~Anchordoqui, I.~Antoniadis and D.~Lust, \emph{{S-dual Quintessence, the Swampland, and the DESI DR2 Results}},  \href{https://arxiv.org/abs/2503.19428}{{\ttfamily 2503.19428}}.

\bibitem{Martin:2000xs}
J.~Martin and R.H.~Brandenberger, \emph{{The TransPlanckian problem of inflationary cosmology}}, \href{https://doi.org/10.1103/PhysRevD.63.123501}{\emph{Phys. Rev. D} {\bfseries 63} (2001) 123501} [\href{https://arxiv.org/abs/hep-th/0005209}{{\ttfamily hep-th/0005209}}].

\bibitem{Rob2000wr}
R.H.~Brandenberger and J.~Martin, \emph{{The Robustness of inflation to changes in superPlanck scale physics}}, \href{https://doi.org/10.1142/S0217732301004170}{\emph{Mod. Phys. Lett. A} {\bfseries 16} (2001) 999} [\href{https://arxiv.org/abs/astro-ph/0005432}{{\ttfamily astro-ph/0005432}}].

\bibitem{Rob2012}
R.H.~Brandenberger and J.~Martin, \emph{{Trans-Planckian Issues for Inflationary Cosmology}}, \href{https://doi.org/10.1088/0264-9381/30/11/113001}{\emph{Class. Quant. Grav.} {\bfseries 30} (2013) 113001} [\href{https://arxiv.org/abs/1211.6753}{{\ttfamily 1211.6753}}].

\bibitem{Agrawal:2018own}
P.~Agrawal, G.~Obied, P.J.~Steinhardt and C.~Vafa, \emph{{On the Cosmological Implications of the String Swampland}}, \href{https://doi.org/10.1016/j.physletb.2018.07.040}{\emph{Phys. Lett. B} {\bfseries 784} (2018) 271} [\href{https://arxiv.org/abs/1806.09718}{{\ttfamily 1806.09718}}].

\bibitem{Heisenberg:2018yae}
L.~Heisenberg, M.~Bartelmann, R.~Brandenberger and A.~Refregier, \emph{{Dark Energy in the Swampland}}, \href{https://doi.org/10.1103/PhysRevD.98.123502}{\emph{Phys. Rev. D} {\bfseries 98} (2018) 123502} [\href{https://arxiv.org/abs/1808.02877}{{\ttfamily 1808.02877}}].

\bibitem{Li:2019ipk}
H.-H.~Li, G.~Ye, Y.~Cai and Y.-S.~Piao, \emph{{Trans-Planckian censorship of multistage inflation and dark energy}}, \href{https://doi.org/10.1103/PhysRevD.101.063527}{\emph{Phys. Rev. D} {\bfseries 101} (2020) 063527} [\href{https://arxiv.org/abs/1911.06148}{{\ttfamily 1911.06148}}].

\bibitem{Cicoli:2020cfj}
M.~Cicoli, G.~Dibitetto and F.G.~Pedro, \emph{{New accelerating solutions in late-time cosmology}}, \href{https://doi.org/10.1103/PhysRevD.101.103524}{\emph{Phys. Rev. D} {\bfseries 101} (2020) 103524} [\href{https://arxiv.org/abs/2002.02695}{{\ttfamily 2002.02695}}].

\bibitem{DiValentino:2025sru}
E.~Di~Valentino et~al., \emph{{The CosmoVerse White Paper: Addressing observational tensions in cosmology with systematics and fundamental physics}},  \href{https://arxiv.org/abs/2504.01669}{{\ttfamily 2504.01669}}.

\bibitem{Copeland:2006wr}
E.J.~Copeland, M.~Sami and S.~Tsujikawa, \emph{{Dynamics of dark energy}}, \href{https://doi.org/10.1142/S021827180600942X}{\emph{Int. J. Mod. Phys. D} {\bfseries 15} (2006) 1753} [\href{https://arxiv.org/abs/hep-th/0603057}{{\ttfamily hep-th/0603057}}].

\bibitem{Chevallier:2000qy}
M.~Chevallier and D.~Polarski, \emph{{Accelerating universes with scaling dark matter}}, \href{https://doi.org/10.1142/S0218271801000822}{\emph{Int. J. Mod. Phys. D} {\bfseries 10} (2001) 213} [\href{https://arxiv.org/abs/gr-qc/0009008}{{\ttfamily gr-qc/0009008}}].

\bibitem{Linder:2002et}
E.V.~Linder, \emph{{Exploring the expansion history of the universe}}, \href{https://doi.org/10.1103/PhysRevLett.90.091301}{\emph{Phys. Rev. Lett.} {\bfseries 90} (2003) 091301} [\href{https://arxiv.org/abs/astro-ph/0208512}{{\ttfamily astro-ph/0208512}}].

\bibitem{Planck:2018vyg}
{\scshape Planck} collaboration, \emph{{Planck 2018 results. VI. Cosmological parameters}}, \href{https://doi.org/10.1051/0004-6361/201833910}{\emph{Astron. Astrophys.} {\bfseries 641} (2020) A6} [\href{https://arxiv.org/abs/1807.06209}{{\ttfamily 1807.06209}}].

\bibitem{DESI:2024hhd}
{\scshape DESI} collaboration, \emph{{DESI 2024 VII: Cosmological Constraints from the Full-Shape Modeling of Clustering Measurements}},  \href{https://arxiv.org/abs/2411.12022}{{\ttfamily 2411.12022}}.

\bibitem{Gu:2025xie}
G.~Gu et~al., \emph{{Dynamical Dark Energy in light of the DESI DR2 Baryonic Acoustic Oscillations Measurements}},  \href{https://arxiv.org/abs/2504.06118}{{\ttfamily 2504.06118}}.

\bibitem{Ratra:1987rm}
B.~Ratra and P.J.E.~Peebles, \emph{{Cosmological Consequences of a Rolling Homogeneous Scalar Field}}, \href{https://doi.org/10.1103/PhysRevD.37.3406}{\emph{Phys. Rev. D} {\bfseries 37} (1988) 3406}.

\bibitem{Caldwell:1999ew}
R.R.~Caldwell, \emph{{A Phantom menace?}}, \href{https://doi.org/10.1016/S0370-2693(02)02589-3}{\emph{Phys. Lett. B} {\bfseries 545} (2002) 23} [\href{https://arxiv.org/abs/astro-ph/9908168}{{\ttfamily astro-ph/9908168}}].

\bibitem{Caldwell:2003vq}
R.R.~Caldwell, M.~Kamionkowski and N.N.~Weinberg, \emph{{Phantom energy and cosmic doomsday}}, \href{https://doi.org/10.1103/PhysRevLett.91.071301}{\emph{Phys. Rev. Lett.} {\bfseries 91} (2003) 071301} [\href{https://arxiv.org/abs/astro-ph/0302506}{{\ttfamily astro-ph/0302506}}].

\bibitem{Cline:2003gs}
J.M.~Cline, S.~Jeon and G.D.~Moore, \emph{{The Phantom menaced: Constraints on low-energy effective ghosts}}, \href{https://doi.org/10.1103/PhysRevD.70.043543}{\emph{Phys. Rev. D} {\bfseries 70} (2004) 043543} [\href{https://arxiv.org/abs/hep-ph/0311312}{{\ttfamily hep-ph/0311312}}].

\bibitem{Chiba:1999ka}
T.~Chiba, T.~Okabe and M.~Yamaguchi, \emph{{Kinetically driven quintessence}}, \href{https://doi.org/10.1103/PhysRevD.62.023511}{\emph{Phys. Rev. D} {\bfseries 62} (2000) 023511} [\href{https://arxiv.org/abs/astro-ph/9912463}{{\ttfamily astro-ph/9912463}}].

\bibitem{Armendariz-Picon:2000nqq}
C.~Armendariz-Picon, V.F.~Mukhanov and P.J.~Steinhardt, \emph{{A Dynamical solution to the problem of a small cosmological constant and late time cosmic acceleration}}, \href{https://doi.org/10.1103/PhysRevLett.85.4438}{\emph{Phys. Rev. Lett.} {\bfseries 85} (2000) 4438} [\href{https://arxiv.org/abs/astro-ph/0004134}{{\ttfamily astro-ph/0004134}}].

\bibitem{Vikman:2004dc}
A.~Vikman, \emph{{Can dark energy evolve to the phantom?}}, \href{https://doi.org/10.1103/PhysRevD.71.023515}{\emph{Phys. Rev. D} {\bfseries 71} (2005) 023515} [\href{https://arxiv.org/abs/astro-ph/0407107}{{\ttfamily astro-ph/0407107}}].

\bibitem{Xia:2007km}
J.-Q.~Xia, Y.-F.~Cai, T.-T.~Qiu, G.-B.~Zhao and X.~Zhang, \emph{{Constraints on the Sound Speed of Dynamical Dark Energy}}, \href{https://doi.org/10.1142/S0218271808012784}{\emph{Int. J. Mod. Phys. D} {\bfseries 17} (2008) 1229} [\href{https://arxiv.org/abs/astro-ph/0703202}{{\ttfamily astro-ph/0703202}}].

\bibitem{Vafa:2005ui}
C.~Vafa, \emph{{The String landscape and the swampland}},  \href{https://arxiv.org/abs/hep-th/0509212}{{\ttfamily hep-th/0509212}}.

\bibitem{vanBeest2021}
M.~van Beest, J.~Calder\'on-Infante, D.~Mirfendereski and I.~Valenzuela, \emph{{Lectures on the Swampland Program in String Compactifications}}, \href{https://doi.org/10.1016/j.physrep.2022.09.002}{\emph{Phys. Rept.} {\bfseries 989} (2022) 1} [\href{https://arxiv.org/abs/2102.01111}{{\ttfamily 2102.01111}}].

\bibitem{Agmon2022}
N.B.~Agmon, A.~Bedroya, M.J.~Kang and C.~Vafa, \emph{{Lectures on the string landscape and the Swampland}},  \href{https://arxiv.org/abs/2212.06187}{{\ttfamily 2212.06187}}.

\bibitem{vafa2020}
M.~Montero and C.~Vafa, \emph{{Cobordism Conjecture, Anomalies, and the String Lamppost Principle}}, \href{https://doi.org/10.1007/JHEP01(2021)063}{\emph{JHEP} {\bfseries 01} (2021) 063} [\href{https://arxiv.org/abs/2008.11729}{{\ttfamily 2008.11729}}].

\bibitem{Rob1990}
R.H.~Brandenberger, R.~Laflamme and M.~Mijic, \emph{{Classical Perturbations From Decoherence of Quantum Fluctuations in the Inflationary Universe}}, \href{https://doi.org/10.1142/S0217732390002651}{\emph{Mod. Phys. Lett. A} {\bfseries 5} (1990) 2311}.

\bibitem{Kiefer2008}
C.~Kiefer and D.~Polarski, \emph{{Why do cosmological perturbations look classical to us?}}, \href{https://doi.org/10.1166/asl.2009.1023}{\emph{Adv. Sci. Lett.} {\bfseries 2} (2009) 164} [\href{https://arxiv.org/abs/0810.0087}{{\ttfamily 0810.0087}}].

\bibitem{Vafa2024}
C.~Vafa, \emph{{Swamplandish Unification of the Dark Sector}},  \href{https://arxiv.org/abs/2402.00981}{{\ttfamily 2402.00981}}.

\bibitem{Vafa2025}
C.~Vafa, \emph{{On the origin and fate of our universe}}, \href{https://doi.org/10.1007/s10714-025-03353-w}{\emph{Gen. Rel. Grav.} {\bfseries 57} (2025) 19} [\href{https://arxiv.org/abs/2501.00966}{{\ttfamily 2501.00966}}].

\bibitem{Lust2022}
S.~L\"ust, C.~Vafa, M.~Wiesner and K.~Xu, \emph{{Holography and the KKLT scenario}}, \href{https://doi.org/10.1007/JHEP10(2022)188}{\emph{JHEP} {\bfseries 10} (2022) 188} [\href{https://arxiv.org/abs/2204.07171}{{\ttfamily 2204.07171}}].

\bibitem{Ooguri2006}
H.~Ooguri and C.~Vafa, \emph{{On the Geometry of the String Landscape and the Swampland}}, \href{https://doi.org/10.1016/j.nuclphysb.2006.10.033}{\emph{Nucl. Phys. B} {\bfseries 766} (2007) 21} [\href{https://arxiv.org/abs/hep-th/0605264}{{\ttfamily hep-th/0605264}}].

\bibitem{ds2018}
G.~Obied, H.~Ooguri, L.~Spodyneiko and C.~Vafa, \emph{{De Sitter Space and the Swampland}},  \href{https://arxiv.org/abs/1806.08362}{{\ttfamily 1806.08362}}.

\bibitem{Scalisi2018}
M.~Scalisi and I.~Valenzuela, \emph{{Swampland distance conjecture, inflation and $\alpha$-attractors}}, \href{https://doi.org/10.1007/JHEP08(2019)160}{\emph{JHEP} {\bfseries 08} (2019) 160} [\href{https://arxiv.org/abs/1812.07558}{{\ttfamily 1812.07558}}].

\bibitem{Bedroya2022}
A.~Bedroya, \emph{{Holographic origin of TCC and the distance conjecture}}, \href{https://doi.org/10.1007/JHEP06(2024)016}{\emph{JHEP} {\bfseries 06} (2024) 016} [\href{https://arxiv.org/abs/2211.09128}{{\ttfamily 2211.09128}}].

\bibitem{Bedroya2024}
A.~Bedroya, Q.~Lu and P.~Steinhardt, \emph{{TCC in the interior of moduli space and its implications for the string landscape and cosmology}},  \href{https://arxiv.org/abs/2407.08793}{{\ttfamily 2407.08793}}.

\bibitem{LiddleLyth2000}
A.R.~Liddle and D.H.~Lyth, \emph{Cosmological Inflation and Large-Scale Structure}, Cambridge University Press (2000).

\bibitem{Yang:2018qmz}
W.~Yang, S.~Pan, E.~Di~Valentino, E.N.~Saridakis and S.~Chakraborty, \emph{{Observational constraints on one-parameter dynamical dark-energy parametrizations and the $H_0$ tension}}, \href{https://doi.org/10.1103/PhysRevD.99.043543}{\emph{Phys. Rev. D} {\bfseries 99} (2019) 043543} [\href{https://arxiv.org/abs/1810.05141}{{\ttfamily 1810.05141}}].

\bibitem{Pan:2019gop}
S.~Pan, W.~Yang, E.~Di~Valentino, E.N.~Saridakis and S.~Chakraborty, \emph{{Interacting scenarios with dynamical dark energy: Observational constraints and alleviation of the $H_0$ tension}}, \href{https://doi.org/10.1103/PhysRevD.100.103520}{\emph{Phys. Rev. D} {\bfseries 100} (2019) 103520} [\href{https://arxiv.org/abs/1907.07540}{{\ttfamily 1907.07540}}].

\bibitem{Colgain:2021pmf}
E.O.~Colg\'ain, M.M.~Sheikh-Jabbari and L.~Yin, \emph{{Can dark energy be dynamical?}}, \href{https://doi.org/10.1103/PhysRevD.104.023510}{\emph{Phys. Rev. D} {\bfseries 104} (2021) 023510} [\href{https://arxiv.org/abs/2104.01930}{{\ttfamily 2104.01930}}].

\bibitem{Giare:2024gpk}
W.~Giar\`e, M.~Najafi, S.~Pan, E.~Di~Valentino and J.T.~Firouzjaee, \emph{{Robust preference for Dynamical Dark Energy in DESI BAO and SN measurements}}, \href{https://doi.org/10.1088/1475-7516/2024/10/035}{\emph{JCAP} {\bfseries 10} (2024) 035} [\href{https://arxiv.org/abs/2407.16689}{{\ttfamily 2407.16689}}].

\bibitem{Barboza:2008rh}
E.M.~Barboza, Jr. and J.S.~Alcaniz, \emph{{A parametric model for dark energy}}, \href{https://doi.org/10.1016/j.physletb.2008.08.012}{\emph{Phys. Lett. B} {\bfseries 666} (2008) 415} [\href{https://arxiv.org/abs/0805.1713}{{\ttfamily 0805.1713}}].

\bibitem{Dimakis:2016mip}
N.~Dimakis, A.~Karagiorgos, A.~Zampeli, A.~Paliathanasis, T.~Christodoulakis and P.A.~Terzis, \emph{{General Analytic Solutions of Scalar Field Cosmology with Arbitrary Potential}}, \href{https://doi.org/10.1103/PhysRevD.93.123518}{\emph{Phys. Rev. D} {\bfseries 93} (2016) 123518} [\href{https://arxiv.org/abs/1604.05168}{{\ttfamily 1604.05168}}].

\bibitem{Pan:2019brc}
S.~Pan, W.~Yang and A.~Paliathanasis, \emph{{Imprints of an extended Chevallier\textendash{}Polarski\textendash{}Linder parametrization on the large scale of our universe}}, \href{https://doi.org/10.1140/epjc/s10052-020-7832-y}{\emph{Eur. Phys. J. C} {\bfseries 80} (2020) 274} [\href{https://arxiv.org/abs/1902.07108}{{\ttfamily 1902.07108}}].

\bibitem{Efstathiou:1999tm}
G.~Efstathiou, \emph{{Constraining the equation of state of the universe from distant type Ia supernovae and cosmic microwave background anisotropies}}, \href{https://doi.org/10.1046/j.1365-8711.1999.02997.x}{\emph{Mon. Not. Roy. Astron. Soc.} {\bfseries 310} (1999) 842} [\href{https://arxiv.org/abs/astro-ph/9904356}{{\ttfamily astro-ph/9904356}}].

\bibitem{Jassal:2005qc}
H.K.~Jassal, J.S.~Bagla and T.~Padmanabhan, \emph{{Observational constraints on low redshift evolution of dark energy: How consistent are different observations?}}, \href{https://doi.org/10.1103/PhysRevD.72.103503}{\emph{Phys. Rev. D} {\bfseries 72} (2005) 103503} [\href{https://arxiv.org/abs/astro-ph/0506748}{{\ttfamily astro-ph/0506748}}].

\bibitem{Caldwell:2005tm}
R.R.~Caldwell and E.V.~Linder, \emph{{The Limits of quintessence}}, \href{https://doi.org/10.1103/PhysRevLett.95.141301}{\emph{Phys. Rev. Lett.} {\bfseries 95} (2005) 141301} [\href{https://arxiv.org/abs/astro-ph/0505494}{{\ttfamily astro-ph/0505494}}].

\bibitem{Wolf:2023uno}
W.J.~Wolf and P.G.~Ferreira, \emph{{Underdetermination of dark energy}}, \href{https://doi.org/10.1103/PhysRevD.108.103519}{\emph{Phys. Rev. D} {\bfseries 108} (2023) 103519} [\href{https://arxiv.org/abs/2310.07482}{{\ttfamily 2310.07482}}].

\bibitem{Cai2015}
Y.-F.~Cai, S.~Capozziello, M.~De~Laurentis and E.N.~Saridakis, \emph{{f(T) teleparallel gravity and cosmology}}, \href{https://doi.org/10.1088/0034-4885/79/10/106901}{\emph{Rept. Prog. Phys.} {\bfseries 79} (2016) 106901} [\href{https://arxiv.org/abs/1511.07586}{{\ttfamily 1511.07586}}].

\bibitem{Bahamonde:2021gfp}
S.~Bahamonde, K.F.~Dialektopoulos, C.~Escamilla-Rivera, G.~Farrugia, V.~Gakis, M.~Hendry et~al., \emph{{Teleparallel gravity: from theory to cosmology}}, \href{https://doi.org/10.1088/1361-6633/ac9cef}{\emph{Rept. Prog. Phys.} {\bfseries 86} (2023) 026901} [\href{https://arxiv.org/abs/2106.13793}{{\ttfamily 2106.13793}}].

\bibitem{BeltranJimenez:2019tme}
J.~Beltr\'an~Jim\'enez, L.~Heisenberg, T.S.~Koivisto and S.~Pekar, \emph{{Cosmology in $f(Q)$ geometry}}, \href{https://doi.org/10.1103/PhysRevD.101.103507}{\emph{Phys. Rev. D} {\bfseries 101} (2020) 103507} [\href{https://arxiv.org/abs/1906.10027}{{\ttfamily 1906.10027}}].

\bibitem{Heisenberg:2018vsk}
L.~Heisenberg, \emph{{A systematic approach to generalisations of General Relativity and their cosmological implications}}, \href{https://doi.org/10.1016/j.physrep.2018.11.006}{\emph{Phys. Rept.} {\bfseries 796} (2019) 1} [\href{https://arxiv.org/abs/1807.01725}{{\ttfamily 1807.01725}}].

\bibitem{BeltranJimenez:2019esp}
J.~Beltr\'an~Jim\'enez, L.~Heisenberg and T.S.~Koivisto, \emph{{The Geometrical Trinity of Gravity}}, \href{https://doi.org/10.3390/universe5070173}{\emph{Universe} {\bfseries 5} (2019) 173} [\href{https://arxiv.org/abs/1903.06830}{{\ttfamily 1903.06830}}].

\bibitem{Wu:2024vcr}
C.~Wu, X.~Ren, Y.~Yang, Y.-M.~Hu and E.N.~Saridakis, \emph{{Background-dependent and classical correspondences between $f(Q)$ and $f(T)$ gravity}},  \href{https://arxiv.org/abs/2412.01104}{{\ttfamily 2412.01104}}.

\bibitem{Basilakos:2025olm}
S.~Basilakos, A.~Paliathanasis and E.N.~Saridakis, \emph{{Equivalence of $f(Q)$ cosmology with quintom-like scenario: the phantom field as effective realization of the non-trivial connection}},  \href{https://arxiv.org/abs/2503.19864}{{\ttfamily 2503.19864}}.

\bibitem{Linder_2010}
E.V.~Linder, \emph{Einstein’s other gravity and the acceleration of the universe}, \href{https://doi.org/10.1103/physrevd.81.127301}{\emph{Physical Review D} {\bfseries 81} (2010) }.

\bibitem{Nesseris:2013jea}
S.~Nesseris, S.~Basilakos, E.N.~Saridakis and L.~Perivolaropoulos, \emph{{Viable $f(T)$ models are practically indistinguishable from $\Lambda$CDM}}, \href{https://doi.org/10.1103/PhysRevD.88.103010}{\emph{Phys. Rev. D} {\bfseries 88} (2013) 103010} [\href{https://arxiv.org/abs/1308.6142}{{\ttfamily 1308.6142}}].

\bibitem{Bamba_2011}
K.~Bamba, C.-Q.~Geng, C.-C.~Lee and L.-W.~Luo, \emph{Equation of state for dark energy in {f(T)} gravity}, \href{https://doi.org/10.1088/1475-7516/2011/01/021}{\emph{Journal of Cosmology and Astroparticle Physics} {\bfseries 2011} (2011) 021–021}.

\bibitem{Lovelock1971}
D.~Lovelock, \emph{{The Einstein tensor and its generalizations}}, \href{https://doi.org/10.1063/1.1665613}{\emph{J. Math. Phys.} {\bfseries 12} (1971) 498}.

\bibitem{Hu2023}
Y.-M.~Hu, Y.~Zhao, X.~Ren, B.~Wang, E.N.~Saridakis and Y.-F.~Cai, \emph{{The effective field theory approach to the strong coupling issue in f(T) gravity}}, \href{https://doi.org/10.1088/1475-7516/2023/07/060}{\emph{JCAP} {\bfseries 07} (2023) 060} [\href{https://arxiv.org/abs/2302.03545}{{\ttfamily 2302.03545}}].

\bibitem{Hu2023xcf}
Y.-M.~Hu, Y.~Yu, Y.-F.~Cai and X.~Gao, \emph{{The effective field theory approach to the strong coupling issue in f(T) gravity with a non-minimally coupled scalar field}}, \href{https://doi.org/10.1088/1475-7516/2024/03/025}{\emph{JCAP} {\bfseries 03} (2024) 025} [\href{https://arxiv.org/abs/2311.12645}{{\ttfamily 2311.12645}}].

\bibitem{Maggiore2014}
M.~Maggiore and M.~Mancarella, \emph{{Nonlocal gravity and dark energy}}, \href{https://doi.org/10.1103/PhysRevD.90.023005}{\emph{Phys. Rev. D} {\bfseries 90} (2014) 023005} [\href{https://arxiv.org/abs/1402.0448}{{\ttfamily 1402.0448}}].

\bibitem{Cai2009}
Y.-F.~Cai and E.N.~Saridakis, \emph{{Non-singular cosmology in a model of non-relativistic gravity}}, \href{https://doi.org/10.1088/1475-7516/2009/10/020}{\emph{JCAP} {\bfseries 10} (2009) 020} [\href{https://arxiv.org/abs/0906.1789}{{\ttfamily 0906.1789}}].

\bibitem{Horava2009}
P.~Horava, \emph{{Quantum Gravity at a Lifshitz Point}}, \href{https://doi.org/10.1103/PhysRevD.79.084008}{\emph{Phys. Rev. D} {\bfseries 79} (2009) 084008} [\href{https://arxiv.org/abs/0901.3775}{{\ttfamily 0901.3775}}].

\bibitem{wang2016}
B.~Wang, E.~Abdalla, F.~Atrio-Barandela and D.~Pavon, \emph{{Dark Matter and Dark Energy Interactions: Theoretical Challenges, Cosmological Implications and Observational Signatures}}, \href{https://doi.org/10.1088/0034-4885/79/9/096901}{\emph{Rept. Prog. Phys.} {\bfseries 79} (2016) 096901} [\href{https://arxiv.org/abs/1603.08299}{{\ttfamily 1603.08299}}].

\bibitem{Bilic2001}
N.~Bilic, G.B.~Tupper and R.D.~Viollier, \emph{{Unification of dark matter and dark energy: The Inhomogeneous Chaplygin gas}}, \href{https://doi.org/10.1016/S0370-2693(02)01716-1}{\emph{Phys. Lett. B} {\bfseries 535} (2002) 17} [\href{https://arxiv.org/abs/astro-ph/0111325}{{\ttfamily astro-ph/0111325}}].

\bibitem{Bento2002}
M.C.~Bento, O.~Bertolami and A.A.~Sen, \emph{{Generalized Chaplygin gas, accelerated expansion and dark energy matter unification}}, \href{https://doi.org/10.1103/PhysRevD.66.043507}{\emph{Phys. Rev. D} {\bfseries 66} (2002) 043507} [\href{https://arxiv.org/abs/gr-qc/0202064}{{\ttfamily gr-qc/0202064}}].

\bibitem{Manous2017}
G.~Koutsoumbas, K.~Ntrekis, E.~Papantonopoulos and E.N.~Saridakis, \emph{{Unification of Dark Matter - Dark Energy in Generalized Galileon Theories}}, \href{https://doi.org/10.1088/1475-7516/2018/02/003}{\emph{JCAP} {\bfseries 02} (2018) 003} [\href{https://arxiv.org/abs/1704.08640}{{\ttfamily 1704.08640}}].

\bibitem{Vafa2024fpx}
C.~Vafa, \emph{{Swamplandish Unification of the Dark Sector}},  \href{https://arxiv.org/abs/2402.00981}{{\ttfamily 2402.00981}}.

\end{thebibliography}\endgroup

\end{document}